\documentclass{aa}  

\usepackage{graphicx}
\usepackage{txfonts}

\usepackage{lscape}
\usepackage{afterpage}
\usepackage{epigraph}
\usepackage{capt-of}

\usepackage[title]{appendix}
\usepackage{natbib}
\usepackage{nccmath}

\usepackage[flushleft]{threeparttable}

\begin{document}

   \title{Probing \textit{Kepler's} hottest small planets via homogeneous search and analysis of optical secondary eclipses and phase variations}


   \author{Vikash Singh
          \inst{1,2}\thanks{vikash.singh@inaf.it}
          ,
          A. S. Bonomo\inst{3}
          ,
          G. Scandariato\inst{2}
          ,
          N. Cibrario\inst{4}
          ,
          D. Barbato\inst{5,3}
          ,
          L. Fossati\inst{6}
          ,
          I. Pagano\inst{2}
          , 
          A. Sozzetti\inst{3}
          }

   \institute{\inst{1} 
                Dip. di Fisica e Astronomia “Ettore Majorana”, Università di Catania, 
                Via S.Sofia 64, I-95123, Catania, Italy\\
             \inst{2} 
                INAF – Osservatorio Astrofisico di Catania, Via S.Sofia 78, I-95123, Catania, Italy
                \thanks{}\\
             \inst{3}
                INAF – Osservatorio Astrofisico di Torino, Via Osservatorio, 20, 10025 Pino Torinese TO, Italy\\
            \inst{4}
                Dipartimento di Fisica, Universit\`a degli Studi di Torino, via Pietro Giuria 1, I-10125, Torino, Italy\\
            \inst{5}
                Observatoire de Gen\`{e}ve, Universit\'{e} de Gen\`{e}ve, 51 Chemin des Maillettes, CH-1290 Sauverny, Switzerland\\
            \inst{6}
                Space Research Institute, Austrian Academy of Sciences, Schmiedlstr. 6, 8042 Graz, Austria
             }

   \date{Received July 27, 2020; accepted November 8, 2021}

 
  \abstract
   {High-precision photometry can lead to the detection of secondary eclipses and phase variations of highly irradiated planets.}
   {We performed a homogeneous search and analysis of optical occultations and phase variations of the most favorable ultra-short-period (USP) ($P<1$~d) sub-Neptunes ($R_{p}<4 R_{\oplus}$), observed by $\textit{Kepler}$ and K2, with the aim to better understand their nature.}
   {We first selected 16 $\textit{Kepler}$ and K2 USP sub-Neptunes based on the expected occultation signal. We filtered out stellar variability in the $\textit{Kepler}$ light curves, using a sliding linear fitting and, when required, a more sophisticated approach based on a Gaussian process regression. In the case of the detection of secondary eclipse or phase variation with a confidence level higher than $2\sigma$, we simultaneously modeled the primary transit, secondary eclipse, and phase variations in a Bayesian framework, by using information from previous studies and knowledge of the Gaia parallaxes. We further derived constraints on the geometric albedo as a function of the planet's brightness temperature.}
   {We confirm the optical secondary eclipses for Kepler-10b ($13\sigma$), Kepler-78b ($9.5\sigma$),  and K2-141b ($6.9\sigma$), with marginal evidence for K2-312b ($2.2\sigma$). We report new detections for K2-106b ($3.3\sigma$), K2-131b (3.2$\sigma$), Kepler-407b ($3.0\sigma$), and hints for K2-229b (2.5$\sigma$). For all targets, with the exception of K2-229b and K2-312b, we also find phase curve variations with a confidence level higher than $2\sigma$.}
   {Two USP planets, namely Kepler-10b and Kepler-78b, show non-negligible nightside emission. This questions the scenario of magma-ocean worlds with inefficient heat redistribution to the nightside for both planets. Due to the youth of the Kepler-78 system and the small planetary orbital separation, the planet may still retain a collisional secondary atmosphere capable of conducting heat from the day to the nightside. Instead, the presence of an outgassing magma ocean on the dayside and the low high-energy irradiation of the old host star may have enabled Kepler-10b to build up and retain a recently formed collisional secondary atmosphere. The magma-world scenario may instead apply to K2-141b and K2-131b. 
   } 


    \titlerunning{Secondary eclipse and phase variations of Kepler USPs}
    \authorrunning{Singh et al.,}
   \maketitle
   
%

\section{Introduction}

Ultra-short-period (USP, $P < 1$~d) sub-Neptunes ($R_{p}<4 R_{\oplus}$) are intriguing planets because of the very close proximity to their host stars. They are about as common as hot Jupiters \citep{WinnJoshua2018} and are often members of multi-planet systems \citep{Sanchis-Ojeda2014}. The existence of USP sub-Neptunes is not easily explained by current formation and migration models \citep{2018ApJCarrera}, although it is generally believed that they migrated from outer regions. Indeed, their formation is highly improbable at the present location because of temperatures higher than the dust sublimation radius ($a/R_{\star} \sim 8$ for Sun-like stars; e.g., \citealt{Isella2006A&A}).

So far, a small number of USP planets have been well characterized thanks to space-based photometry and high-precision radial-velocity follow-up studies, such as Corot-7b \citep{leger2011extreme, 2014MNRASHaywood}, Kepler-10b \citep{batalha2011kepler, 2014ApJDumusque, 2016ApJWeiss}, Kepler-78b \citep{sanchis2013transits, Pepe2013, Howard_2013Nature}, K2-141b \citep{malavolta2018ultra, Barragan2018A&A}, 55 Cancri e \citep{2016NaturDemory55CncE, 2014MNRAS_Nelson55CncE}, K2-229b \citep{Santerne2018Nat}, WASP-47e \citep{2017AJ....153..265Weiss, 2017AJVanderbergW47e}, K2-131b \citep{2017AJDai}, K2-106b \citep{2017AJSinuKoff, Guenther2017A&A}, and K2-312b \citep{2020A&AFrustagli}. Most USP small planets have a rocky composition with varying silicate-to-iron mass fractions \citep{Dai2019} and are generally less massive than $8~M_{\oplus}$. Some of them could be the remnant cores of hot Neptunes or sub-Saturns that lost their primordial H or He atmospheres under the action of the intense stellar irradiation, particularly, while the star was still young and, hence, active (photo-evaporation) \citep{LeCavelier2007A&A, winn2017absence, kubyshkina2018grid}. Depending on the planet's surface gravity and on the timescale and strength of the competing photo-evaporation and surface outgassing or release processes, USP planets may or may not possess an atmospheric envelope \citep{miguel2011, 2013DPSLopezFortney, owen2017evaporation, ito2015, kislyakova2020induction-aa}. Only a collisional atmosphere, with a relatively high molecular weight in the case of USP planets, may be able to retain and redistribute heat, as might be the case of 55\,Cnc\,e \citep{DawsonFabrycky2010ApJ, Winn2011ApJ, Demory2011A&A, 2016MNRASDemory, Dai2019}.

The surface of rocky USP planets may be at least partially covered by magma because, at their equilibrium temperatures (typically higher than $\sim$2200~K), silicates melt \citep{schaefer2004, schaefer2009, schaefer2012}. In this scenario of "magma-ocean worlds," a very high temperature contrast is expected between the continuously illuminated dayside of the planet and its hidden nightside. Indeed, any possible non-collisional atmosphere (i.e., exosphere) of metal vapors above the magma surface would be unable to transfer heat to the nightside \citep{leger2011extreme}. As a result, the nightside flux of a magma-ocean world is expected to be negligible. 

The observation of transiting systems from space-based high-precision photometry as provided by the optical \textit{Kepler} Space Telescope allows us to detect secondary eclipses and phase variations \citep{esteves2013optical, Esteves2015ApJ, Sheets2017AJ, Jansen2018MNRAS}, especially when a large number of orbital phases can be combined as in the case of USP planets. The secondary eclipse depth is a measure of the brightness of the planet and, in the optical band, results from two contributions: the reflection of stellar light from the planetary atmosphere and the planet thermal emission. The latter may in principle be the dominant contribution for USP sub-Neptunes, given the very high irradiation these planets receive from their host stars. With optical data only, it is not possible to disentangle the reflected and thermal contributions \citep{cowan2011statistics}. Nevertheless, non-negligible nightside flux -- as derived from the difference between the occultation depth and the flux level just before (after) the transit ingress (egress) -- can only be due to thermal nightside emission and, hence, indicate the presence of either a mechanism capable of heating the planetary nightside surface or of heat redistribution from the dayside to the nightside. 

In this work, we present a homogeneous search for and analysis of secondary eclipses and phase variations in \textit{Kepler} and K2 USP sub-Neptunes, with the precise aim of better understanding  the nature of these planets. 

\section{Target selection}\label{sec: TargetSelection}

The USP sub-Neptunes considered in our study were selected from the \textit{Kepler} and K2 catalogs of confirmed planets with $P<1$~d, $R_{p} < 4 R_{\oplus}$ and a \textit{Kepler} magnitude of the host star $<14$. The data were collected from the NASA Exoplanet Archive\footnote{https://exoplanetarchive.ipac.caltech.edu/index.html}. \textit{Kepler} is the most precise space-based telescope to date, both in terms of its exquisite photometric precision and the length of photometric datasets (stellar light curves). 

Given the size and distance of the USP planets from their host star, a secondary eclipse can be observed if the signal-to-noise ratio (S/N) is high enough to enable the detection of a flux decrease of at least a few ppm, when the planet passes behind its host star. For each selected USP planet, we computed the expected S/N for the secondary eclipse as:

\begin{equation}\label{eqn:SNR}
S/N = \frac{\delta_{ec}}{\sigma} \sqrt{N_{ep}}
,\end{equation}

\noindent
where $\sigma$ is the photometric precision of the \textit{Kepler} light curve, which is related to the brightness of the host star in the \textit{Kepler} band-pass; $N_{ep}$ is the number of data points during all observed eclipses: the larger the number of planetary orbits observed by \textit{Kepler}, the higher $N_{ep}$ and, thus, the S/N of the occultation. $\delta_{ec}$ is the estimated depth of the secondary eclipse that was computed with Eqs.~\ref{eqn:EclipseDepth}, \ref{eqn:Albedo}, and \ref{eqn:ThermalFlux} (Sect. \ref{sec: Ag_vs_Tday}) by i) using the stellar and planetary parameters from previous studies; and ii) considering a conservative Bond albedo of $A_{B}$=0.5 under the assumption of isotropic scattering by a perfect Lambertian sphere, namely,  $A_{g}=2/3 \cdot A_{B}$ \citep{rowe2006upper}, and no heat redistribution to the nightside (Eq.~\ref{eqn:Dayside} in Sect. \ref{sec: DayNightTemperatures} for $\epsilon=0$). 

We selected only those targets with theoretically computed $S/N \ge 2$ on the secondary eclipse depth (Eq. \ref{eqn:SNR}). The USP planets that meet this criterion are Kepler-10b, Kepler-78b, Kepler-407b, Kepler-1171b, and Kepler-1323b among the Kepler planets, and K2-23e/WASP-47e, K2-96b/HD 3167b, K2-100b, K2-106b, K2-131b, K2-141b, K2-157b, K2-183b, K2-229b, K2-266b, and K2-312b/HD 80653b from the K2 confirmed planets. However, after the analysis of the \textit{Kepler} and K2 light curves, only eight USP sub-Neptunes: mainly Super-Earths, namely, Kepler-10b, Kepler-78b, Kepler-407b, K2-106b, K2-131b, K2-141b, K2-229b, and K2-312b, are found to show optical secondary eclipses with $S/N > 2$ (see Sect.~\ref{sec: results}).

\section{Stellar parameters}\label{sec: stellarparam}

With the availability of DR2 Gaia parallaxes \citep{gaia2018}, we improved the parameters of the stellar hosts of the above-mentioned eight systems with secondary eclipses detected at $S/N > 2$, to impose a prior on the stellar density \citep{2010exop.bookWinn} in the global modeling of transits, secondary eclipses, and phase variations (see Sect.~\ref{sec: modeling}). We re-determined the radius and mass of the host stars by fitting the stellar Spectral Energy Distribution (SED) and using the MESA Isochrones and Stellar Tracks (MIST) \citep{dotter2016,choi2016} with the publicly available \texttt{EXOFASTv2} code \citep{eastman2019}, which makes use of the differential evolution Markov chain Monte Carlo (DE-MCMC, \citealt{TerBraak2006}) Bayesian technique. 
The stellar parameters were simultaneously constrained by the SED and the MIST isochrones, as the SED primarily constrains the stellar radius $R_*$ and effective temperature $T_{eff}$, and a penalty for straying from the MIST evolutionary tracks ensures that the resulting star is physical in nature. 

\par For each star, we fitted the available archival magnitudes from the WISE bands \citep{cutri2014}, Sloan bands from the APASS DR9 \citep{henden2015}, Johnson's B, V, R and 2MASS J, H, K bands from the UCAC4 catalog \citep{zacharias2012}, the \textit{Kepler} \citep{greiss2012} and Gaia \citep{gaia2018} bands. We imposed Gaussian priors on the Gaia DR2 parallax as well as on the stellar effective temperature $T_{eff}$ and metallicity [Fe/H] from the literature values. The DR2 parallax value was first corrected for the systematic offset of $82\pm33$ $\mu$as as reported in \citet{stassun2018}. The prior on the parallax greatly helps constraining the stellar radius and in general improves the accuracy and precision of the stellar parameters. In Table~\ref{tab: Stellar Parameter}, we present the improved parameters of the stars hosting the planets for which we found the secondary eclipse at $S/N > 2$. For Kepler-10, we used the very precise stellar density as computed through asteroseismology \citep{Fogtman-Schultz2014ApJ}. Along with it, we obtained the mass, radius, luminosity, and age from the same source.
The stellar density determination was later used as a prior in the simultaneous fit of the transit, secondary eclipse, and phase variation (see Sect.~\ref{sec: modeling}).

\begin{table*}[h!]
 \centering
    \caption{Stellar parameters from the SED fitting and the MIST evolutionary tracks \citep{eastman2019}.}\label{tab:StellarPars}
  \begin{tabular}{lcccccccccl}\hline\hline
  Star &$M_{\star}$ & $R_{\star}$ & $\rho_{\star}$ & $T_{eff}$ & log g & Metallicity & u1 & u2 \\
   & [$M_{\odot}$]  & [$R_{\odot}$] & [g/cc] & [K] & $log_{10}([cgs])$ & [Fe/H] &  & \\\hline 
   \\
   Kepler-10 & 0.913 $\pm$ 0.022 & 1.065 $\pm$ 0.008 & $1.068 \pm 0.002$ & $5711\substack{+28\\-28}$& $4.309 \substack{+0.063\\-0.090}$ & $-0.150 \substack{+0.040\\-0.040}$ & 0.403(16) & 0.256(10) \\\\
   Kepler-78 & $0.775\substack{+0.038\\-0.033}$ & $0.741\substack{+0.046\\-0.037}$ & $2.67\substack{+0.40\\-0.39}$ & $5119\substack{+44\\-46}$& $4.587 \substack{+0.039\\-0.045}$ & $-0.089 \substack{+0.042\\-0.039}$ & 0.519(11) & 0.178(08) \\\\
   Kepler-407 & $0.967\substack{+0.055\\-0.049}$ & $0.962\substack{+0.037\\-0.029}$ & $1.53\substack{+0.18\\-0.18}$ & $5534\substack{+61\\-60}$& $4.456 \substack{+0.039\\-0.040}$ & $0.279 \substack{+0.065\\-0.065}$ & 0.463(14) & 0.220(08) \\\\
   K2-141 & $0.687\substack{+0.031\\-0.029}$ & $0.668\substack{+0.027\\-0.024}$ & $3.24\substack{+0.33\\-0.30}$ & $4594\substack{+71\\-66}$& $4.625\substack{+0.028\\-0.027}$  & $-0.078 \substack{+0.095\\-0.093}$ & 0.619(10) & 0.106(07) \\\\
   K2-131 & $0.769\substack{+0.033\\-0.029}$ & $0.774\substack{+0.038\\-0.036}$ &  $2.63\substack{+0.37\\-0.33}$ & $5232\substack{+52\\-51}$& $4.581\substack{+0.038\\-0.038}$ & $-0.174 \substack{+0.030\\-0.030}$ & 0.487(12) & 0.201(08) \\\\
   K2-106 & $0.925\substack{+0.049\\-0.042}$ & $0.951\substack{+0.027\\-0.026}$ &  $1.52\substack{+0.15\\-0.13}$ & $5598\substack{+80\\-78}$& $4.449 \substack{+0.031\\-0.029}$ & $0.096 \substack{+0.060\\-0.058}$ & 0.435(16) & 0.237(10) \\\\
   K2-229 & $0.802\substack{+0.033\\-0.028}$ & $0.777\substack{+0.014\\-0.014}$ &  $2.41\substack{+0.15\\-0.14}$ & $5309\substack{+34\\-34}$& $4.561 \substack{+0.021\\-0.020}$ & $-0.091 \substack{+0.020\\-0.020}$ & 0.478(07) & 0.207(05) \\\\
   K2-312 & $1.179\substack{+0.037\\-0.046}$ & $1.221\substack{+0.014\\-0.013}$ & 0.64$\pm$ 0.04& 6022$\pm$72 & 4.34 $\pm$0.02 & 0.29 $\pm$0.03 & 0.371(12) & 0.279(07) \\\\\hline
  \end{tabular}\label{tab: Stellar Parameter}
      \begin{tablenotes}
      \small
      \item For Kepler-10 we adopted the spectroscopic and asteroseismically-determined  parameters \citep{Fogtman-Schultz2014ApJ, 2014ApJDumusque}. \\
      \item The limb darkening coefficients were computed for the \textit{Kepler} passband following \citet{2010A&ASing}.
  \end{tablenotes}
\end{table*}

\section{Light curve analysis}\label{sec: analysis1}

For the selected Kepler targets (Sect.~\ref{sec: TargetSelection}), we downloaded the \textit{Kepler} light curves, both in long-cadence (29.4~min) and short-cadence (58~s) sampling (when available), from the \textit{Kepler} Mikulski Archive for Space Telescopes (MAST)\footnote{http://archive.stsci.edu/}. For the K2 targets, we used the light curves extracted and calibrated by \citet{vanderburg2014technique} and \citet{vanderburg2016planetary}. The light curves were corrected for possible contamination of background stars as reported in the \textit{Kepler} Input Catalogue \citep{2011AJBrown}. 

\subsection{Search for secondary eclipses and phase variations}\label{sec: detection}
\begin{figure}
\centering
\includegraphics[width=\linewidth]{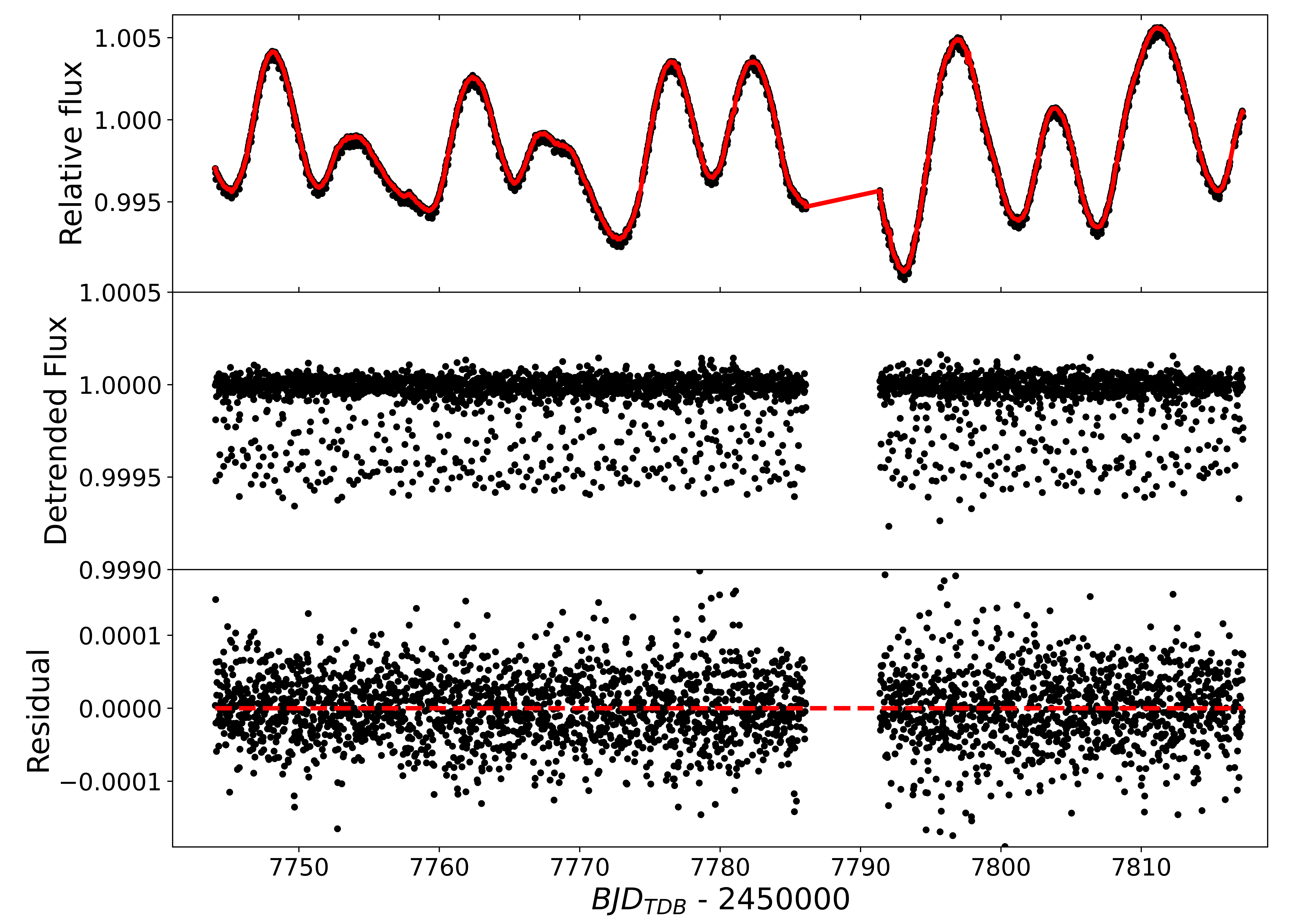}
\hspace{1.0cm}
\caption{Detrending of the stellar variability in the light curve of K2-141 using the sliding linear fitting. Top: K2-141 light curve after removing the transits of planet c; the SLF model is over-plotted in red. Middle: Detrended light curve. Bottom: Residual of the model with the best-fit planetary parameters.}
\label{fig: K2-141_SLF}
\end{figure}

\noindent
From the downloaded light curves we removed possible stellar variability due to the rotational modulation of photospheric active regions, following the method described by \citet{sanchis2013transits}. This is basically a sliding linear fitting (SLF) over the out of primary transit and secondary eclipse data points with a timescale equal to the orbital period (cf. Sect. 3.2 in \citealt{sanchis2013transits} for more details). Since the orbital period of USP planets is considerably shorter than typical stellar rotation signals ($P_{\rm rot} \gtrsim 10$~d), in most cases, this method efficiently filters out stellar variations, while preserving in particular the planet phase variations (as an example, see its application to K2-141 in Fig. \ref{fig: K2-141_SLF}). Indeed, it has enabled the discovery of the secondary eclipse and phase variations of Kepler-78b and K2-141b orbiting active stars \citep{sanchis2013transits, malavolta2018ultra}. In the case of a multiple transiting system, the transits of the planetary companions were removed before the filtering of stellar variability. We then phase-folded the detrended light curves for a visual inspection of the secondary eclipse and phase variations. The targets showing a clear signal of a secondary eclipse or a hint of it were further studied in a Bayesian framework through DE-MCMC analyses \citep{TerBraak2006} of the detrended light curves.

\subsection{Model and analysis of transits, secondary eclipses and phase curves}\label{sec: modeling}
We define the model used for the DE-MCMC analyses as a combination of a transit model, a phase curve, and a secondary eclipse model, following \citet{esteves2013optical} as follows:

\begin{equation}
\label{eqn:model}
F(t) = F_{tr}(t) + F_{ph}(t) + F_{ec}(t).
\end{equation}

\par $F_{tr}$ is the transit model with the formalism of \citet{mandel2002analytic} for a quadratic limb-darkened law. The term for phase variations $F_{ph}$ is defined as \citep[][and references therein]{2011exha.bookPerryman}:

\begin{equation}
F_{ph}(\alpha) = A_{ill} \frac{\sin\alpha + (\pi - \alpha)\cos\alpha}{\pi}
,\end{equation}

\noindent
where $A_{ill}$ is the phase amplitude, $\alpha$ $\in$ [0, $\pi$] is the angle between the star and the observer subtended at the planet with $cos \alpha = -sin(i) sin(2\pi(\phi(t) + 0.25))$, i $\in$ [0, $\pi$/2], and $\phi$ being the orbital inclination and the orbital phase, respectively, with $\phi$=0 at mid-transit. The angle ($\phi$+0.25) corresponds to the position at radial velocity maximum for a circular orbit (quadrature). The phase function is truncated near its peak at $\phi$=0.5 for the entire duration of the secondary eclipse because the full illuminated planet disk is occulted by the star.

\par The function of the secondary eclipse \textit{$F_{ec}$} is made up of three parts: i) the out of secondary eclipse flux which is set to zero; ii) the flux at the eclipse ingress and egress which is derived by calculating the area of the planet being occulted by the star during ingress and egress; and iii) the complete occultation part with depth $\delta_{ec}$. The area of the unocculted planetary disk visible to the observer $A_{ec}$ can thus be computed as

\begin{equation}
\begin{medsize}
  A_{ec}=\left\{
  \begin{array}{@{}ll@{}ll@{}}
    \pi p^{2}, &  1+p \le \zeta(t) \\
    p^{2}(\pi - \alpha_{1}) - \alpha_{2} + \frac{\sqrt{4\zeta(t)^{2} - (1 + \zeta(t)^{2} - p^{2})}}{2}, & 1-p<\zeta(t) \le 1+p \\
    0, &  1-p \ge \zeta(t) \\
  \end{array}\right.
\end{medsize}
,\end{equation} 

where $p$ is the planet's radius and $\zeta$(t) is the projected distance between the stellar and the planetary disk centers, both in unit of stellar radius; $\alpha_{1}$ and $\alpha_{2}$ are the cosine angles given by:

\begin{equation}
\begin{aligned}
\cos \alpha_{1} = \frac{p^{2} + \zeta(t)^{2} - 1}{2p\zeta(t)}, \\
\cos \alpha_{2} = \frac{1+\zeta(t)^{2} - p^{2}}{2\zeta(t)}.
\end{aligned}
\end{equation}

Then, $F_{ec}$ is then defined as:

\begin{equation}
F_{ec} = \delta_{ec}\left(\frac{A_{ec}}{\pi p^{2}} - 1\right)
.\end{equation}

The orbits of USP planets are expected to be tidally locked and, hence, circular and co-rotational because of the very strong tidal effects at the extreme proximity to their host star. By considering circular orbits, the free parameters of our model are the orbital period (P), the epoch of mid-transit ($T_{c}$), the transit duration ($T_{dur}$), the orbital inclination ($i$), the planet-to-stellar radius ratio ($p$), the eclipse depth($\delta_{ec}$), and the amplitude of phase variation ($A_{ill}$). A phase offset term was also explored, but it was found to be consistent with zero in the phase curves with the highest precision (Kepler-10b and Kepler-78b; see Fig. \ref{fig: DEMC-Kepler-offset} for Kepler-10b) and was therefore fixed to zero in our model. We included an additional term $\epsilon_{ref}$ to \textit{F(t)} in Eq. \ref{eqn:model} because even though our light curves were normalized to 1 in relative flux by the sliding linear fitting (Sect.~\ref{sec: detection}), this term allows for possible, albeit very small, changes in the reference level of the out-of-transit flux. The limb-darkening coefficients were fixed to the theoretically computed values for the stellar effective temperature, surface gravity, and metallicity \citep{2010A&ASing}. To derive more accurate and precise transit parameters, we imposed a prior on the stellar density using the values in Table \ref{tab:StellarPars} (Sect.~\ref{sec: stellarparam}). This prior affects all the transit parameters except $T_{c}$ and, at the same time, speeds up the DE-MCMC analyses. The stellar density is indeed related to the other parameters through the relation:  

\begin{equation}
    \rho_{*} = \frac{3\pi}{GP^{2}} \left(\frac{a}{R_{*}}\right)^{3}
    \label{eqn: stellar_density}
,\end{equation}
where 
\begin{equation}
\frac{a}{R_{*}}=\frac{1+p}{\sqrt{1-{\cos}^2{\theta_1} {\sin^2{i}}}}
    \label{eqn: a_rs}
\end{equation}
and $\theta_1=-T_{dur}/(2P)$ is the orbital phase at the transit ingress \citep{Gimenez2006A&A}.

For the analysis of the \textit{Kepler} long-cadence data, the model was oversampled at 1~min sampling and then binned to the long-cadence samples to overcome the well-known issue of light-curve distortions due to long integration times \citep{2010MNRASKipping}.   
For the DE-MCMC analyses, we used a Gaussian likelihood function and the Metropolis-Hastings algorithm to accept or reject a proposal step. We followed the prescriptions given by \citet{eastman2013, eastman2019} about the number of DE-MCMC chains (16, specifically, twice the number of free parameters) and the criteria for the removal of burn-in points and the convergence and proper mixing of the chains. The chains were initialized close to the literature values for the transit parameters and relatively close to the expected theoretical values for $\delta_{ec}$ and $A_{ill}$. 
The median values and the 15.86-84.13\% quantiles of the obtained posterior distributions were respectively taken as the best-fit values and $1\sigma$ uncertainties for each fitted parameter.

\subsection{Modeling with Gaussian processes}

For most of the targets, the SLF technique described in Sect.~4.1 performed well in terms of removing stellar variability by leaving practically negligible correlated noise in the residuals ($\lesssim15\%$ relative to the photometric rms), as estimated following \citet{2006MNRAS.373..231P} and \citet{2012A&A...547A.110B}. 
However, in two cases of particularly active stars, namely K2-131 and K2-229, a significantly higher correlated noise was still present after the SLF filtering, that is, 32$\%$ and 43$\%$ of the residual rms, respectively. This indicates that the SLF was unable to account for short-term activity variations of these two stars and, thus, a more sophisticated approach is needed.

As the correlated noise is expected to be a consequence of active regions co-rotating with the stellar surface, we employed a Gaussian process (GP) regression with a Simple Harmonic Oscillator covariance kernel \citep{2017AJForeman-MackeyCelerite}. We thus modeled the unfiltered light curve simultaneously with the GP regression and the planetary model (Eq.~\ref{eqn:model}). We used the \texttt{celerite2} package  \citep{2017AJForeman-MackeyCelerite, F_mackey2018RNAAS} for the GP implementation. The posterior samples of the free parameters were derived with an MCMC method, using the \texttt{emcee} package \citep{Foreman-Mackey2013PASP}. \footnote{The INAF hotcat cluster \citep{taffoni2020chipp,bertocco2019inaf} was used for the analyses.} The results of the GP hyper-parameters, namely the GP amplitude, the damping time scale and the undamped period, are given in Table~\ref{tab:GP_parameters}; the corresponding transit, secondary eclipse, and phase curve parameters of K2-131b and K2-229b are given in Table \ref{tab:BestfitPars}. The residuals of the best fit of the light curves of K2-131 and K2-229 show no significant correlated noise, proving that for these two cases the GP approach performed better than the SLF.

Considering the GP modeling being computationally demanding and time consuming due to the large number of photometric data points, we employed the SLF filtering for all the targets but K2-131 and K2-229, as the leftover correlated noise in the SLF residuals is insignificant. Nonetheless, we compared the results of the two approaches, SLF vs GP, for K2-141 and obtained fully consistent results (Table \ref{tab: Comparing-SLF_GP}). This points to a similar efficiency of the techniques when the residuals of the SLF do not show high correlated noise.

\begin{table*}[]
    \centering
    \begin{center}
        \caption{Best-fit Gaussian process hyper-parameters for the three systems when applied.}
    \begin{tabular}{||c | c | c | c||} 
     \hline
     Parameters & K2-141b & K2-131b & K2-229b \\ [0.5ex] 
     \hline\hline
     Amplitude of GP [h] (ln[rel. flux]) & $-5.63\substack{+0.30\\-0.20}$ & $-5.49\substack{+0.23\\-0.17}$ & $-4.89\substack{+0.29\\-0.23}$ \\ 
     \hline
     Undamped period [$\theta$] ($\rm{d^{-1}}$) & $0.1292\substack{+0.0048\\-0.0047}$ & $0.1503\substack{+0.009\\-0.010}$ & $0.054\substack{+0.007\\-0.009}$ \\
     \hline
     Damping timescale [l] (ln[d]) & $2.92\substack{+0.61\\-0.41}$ & $1.86\substack{+0.46\\-0.34}$ & $2.23\substack{+0.55\\-0.38}$ \\
     \hline
     
    \end{tabular}\label{tab:GP_parameters}
    \end{center}
   
\end{table*}

 \begin{table*}[h!]  
\begin{center}
\caption{Comparison of the transit, secondary eclipse and the phase curve model parameters of K2-141b with two different detrending techniques, namely the sliding linear fitting and Gaussian processes.}
\begin{tabular}{||c | c | c||}
 \hline
 Model parameters & Sliding linear fitting & Gaussian processes \\ [0.5ex] 
 \hline\hline

 $P$ (days) & 0.2803247(12) & 0.2803246(14) \\
 
 $T_{c}$ ($\rm{BJD_{\rm TDB}}$ - 2450000) & 7744.0716(02) & 7744.0711(02) \\ 
 
 $T_{dur}$ (days) & 0.0401(03) & $0.0401(04)$ \\
 $i$ (degree) & $86.57\substack{+2.23\\-2.69}$ & 86.5$\substack{+2.4\\-3.3}$ \\
 $R_{p}/R_{\star}$ & 0.0204(01) & 0.0204(04) \\
 $\delta_{ec}$ (ppm) & $26.2\substack{+3.6\\-3.7}$ & 27.5(4.4) \\
 $A_{ill}$ (ppm) & $23.5\substack{+3.9\\-4.0}$ & $22.4\substack{+4.7\\-4.8}$ \\[1ex] 
 \hline  

\end{tabular}\label{tab: Comparing-SLF_GP}
\end{center}
\end{table*}

\begin{figure}[h!]
\centering
\includegraphics[width=\linewidth]{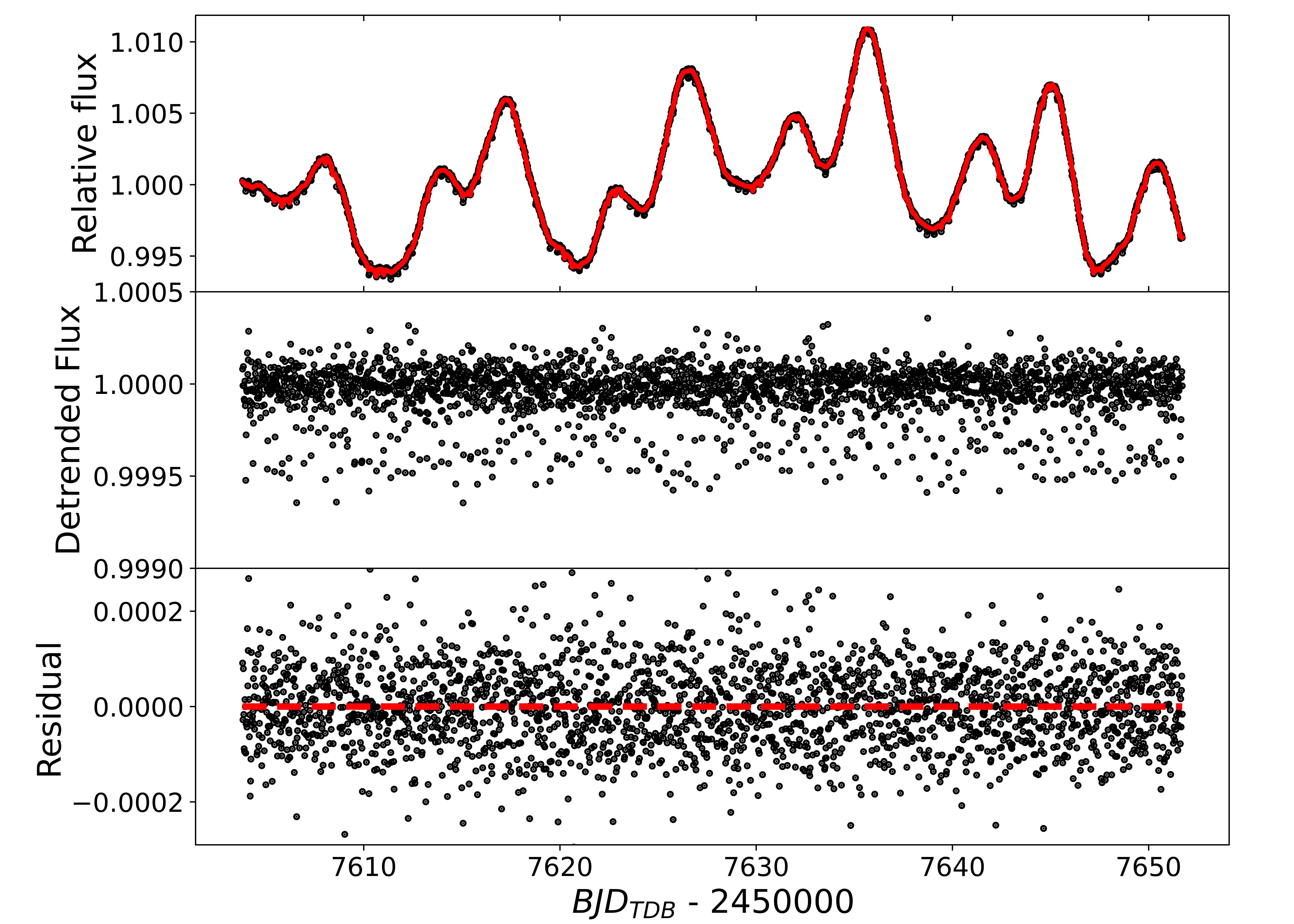}
\hspace{1.0cm}
\caption{Gaussian-process detrending of the stellar variability in the light curve of K2-131. Top: Light curve with the GP model over-plotted in red. Middle: Detrended light curve. Bottom: Best-fit residuals of the combined fit. The GP detrending and model fitting are carried out simultaneously, unlike the SLF detrending.}
\label{fig: K2_131_GP}
\end{figure}

\subsection{Secondary eclipses and constraints on dayside reflection and thermal emission} \label{sec: Ag_vs_Tday}

The optical dayside flux from the secondary eclipse depth is a combination of the reflected light and thermal emission from the planet \citep[e.g.,][]{lopez2007thermal,snellen2009changing}:

\begin{equation}\label{eqn:EclipseDepth}
\delta_{ec} = \delta_{ref} + \delta_{therm}
,\end{equation}  
where $\delta_{ref}$ is the reflected flux and $\delta_{therm}$ represents the dayside thermal emission. The two components can be expressed as:

\begin{equation}\label{eqn:Albedo}
\delta_{ref} = A_{g}\left(\frac{R_{p}}{a}\right)^{2}
,\end{equation}

\begin{equation}\label{eqn:ThermalFlux}
\delta_{therm} = \pi\left(\frac{R_{p}}{R_{\star}}\right)^{2}  \frac{ \int_{\lambda} \frac{2hc^{2}}{\lambda^{5}} \left[ exp \left( \frac{hc}{k_{B}\lambda T_{d}}\right) - 1 \right]^{-1}   \Omega_{\lambda} d\lambda}{\int_{\lambda} S_{\lambda}^{CK} \Omega_{\lambda} d\lambda}
,\end{equation}

\noindent
where $A_{g}$ is the geometric albedo, $a$ the semi-major axis, $h$ is the Planck constant, $k_{B}$ the Boltzmann constant, $c$ the speed of light, $T_{d}$ the planet's dayside brightness temperature. and $S_{\lambda}^{CK}$ is the stellar flux as computed by \citet{Castelli&Kurucz2003IAUS} for the stellar $T_{eff}$, $\log{g,}$ and [Fe/H]; both the planetary and stellar flux are integrated over the \textit{Kepler} passband $\Omega_{\lambda}$\footnote{http://keplergo.arc.nasa.gov/CalibrationResponse.shtml}. In addition, $R_{p}/a$ is derived from the model fit following Eq.~12 in \citet{Gimenez2006A&A}.

In order to estimate the relative fraction of reflection and thermal emission from the occultation depth, Equations \ref{eqn:EclipseDepth}, \ref{eqn:Albedo}, and \ref{eqn:ThermalFlux} are used to compute the geometric albedo as a function of varying dayside brightness temperature.

\subsection{Phase variations and constraints on the nightside emission}

The phase curve conveys information about the flux from the planet dayside as the planet moves along its orbit. The difference between the eclipse depth $\delta_{ec}$ and the amplitude of the phase variation $A_{ill}$ allows us to estimate the nightside flux. Indeed, the flux during the secondary eclipse corresponds to the stellar flux only because the planet is hidden by the star. The flux at the base of the phase curve just before or after transit (phases $T_{1}/T_{4}$ respectively in Fig. \ref{fig:Sec_eclipse}) instead includes the possible contribution from the nightside thermal emission, as well as the flux $f_{cres}^{transit}$ due to the reflection and thermal emission  of the thin bright crescent of the illuminated planet hemisphere. On the other side, just before and after the conjunction (phases $t_{1}/t_{4}$), the dayside hemisphere is not entirely visible because a small fraction of it, say $f_{cres}^{eclipse}$, is hidden from the observer (it is equal in area to the dark crescent that is in view at this epoch).
Therefore, $f_{cres}^{eclipse}$ can be computed as the difference between the maximum of the phase variation at phase 0.5 and the value of the phase curve at the secondary eclipse ingress $t_{1}$ or egress $t_{4}$ as (see Fig.~\ref{fig:Sec_eclipse})

\begin{equation}
f_{cres}^{eclipse} = A_{ill}(max) - A_{ill}(\phi_{t_{1},t_{4}})\,,
\end{equation}

\noindent
where $A_{ill}(\phi_{t_{1}, t_{4}})$ is the phase curve value computed at $t_{1}$ or $t_{4}$. The difference between the flux level $A_{ill}(\phi_{T_{1},T_{4}})$ at the transit ingress (egress) $T_{1}/T_{4}$ and the bottom of the secondary eclipse, that is, the flux from the star alone ($F_{star}$), has two contributions: i) the nightside emission $\delta_{night}$ and ii) the reflection and emission from the crescent of the illuminated hemisphere $f_{cres}^{transit}$. Therefore, as shown in Fig.~\ref{fig:Sec_eclipse}, we can compute the nightside flux $\delta_{night}$~\footnote{$\delta_{night}$ is missing the tiny fraction of the nightside flux which is facing away from the observer. Similarly, $f_{cres}^{eclipse}$ includes that same tiny fraction of the nightside flux, this time facing towards the observer} from the following set of equations:

\begin{figure}[h!]
\centering
\includegraphics[width=\linewidth]{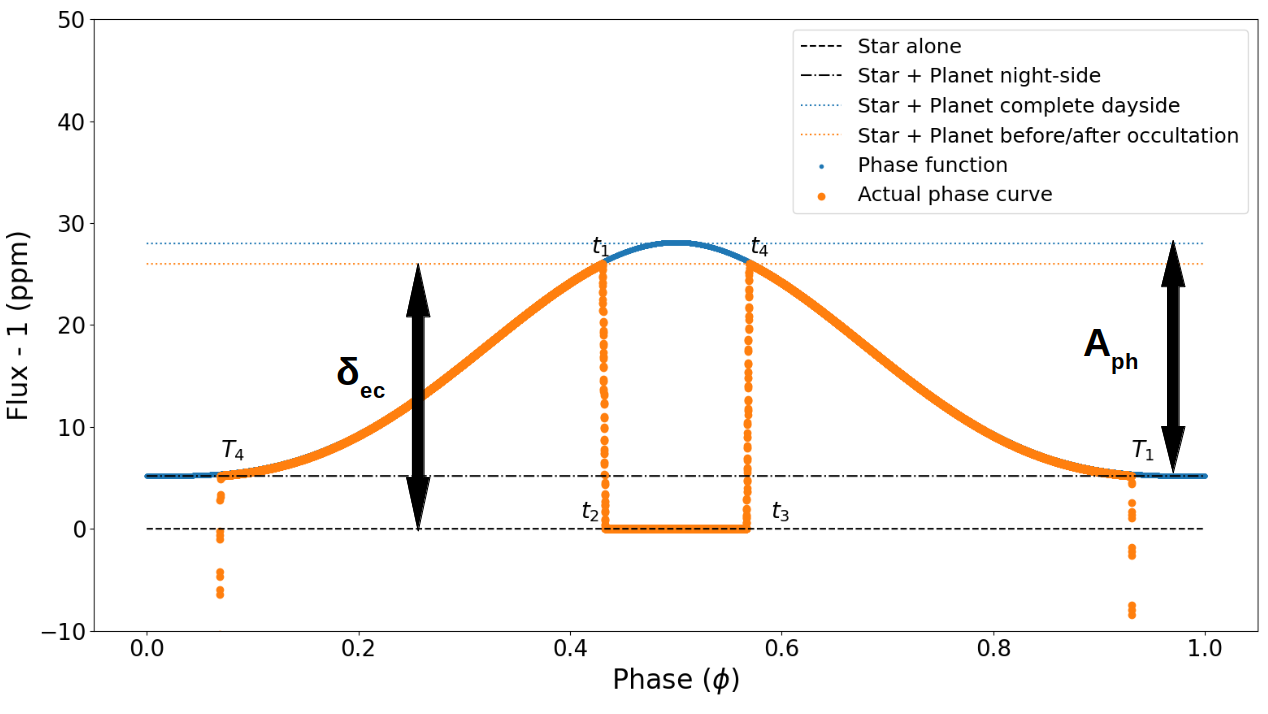}
\hspace{1.0cm}
\caption{Phase curve of a circular orbit around the secondary eclipse showing the difference flux reference levels. The duration of the secondary eclipse is equal to the transit duration. The times $t_1$, $t_2$, $t_3$, and $t_4$ define the ingress and egress moments of the planetary occultation in the same way as $T_1$, $T_2$, $T_3$, $T_4$ for the primary transit. The eclipse depth is actually the difference between the flux references of planet's brightness at $t_1/t_4$ and the stellar brightness at $t_2/t_3$. }
\label{fig:Sec_eclipse}
\end{figure}

\begin{equation}
    \delta_{ec} = A_{ill}(\phi_{t1}, \phi_{t4}) - F_{star}
,\end{equation}
\begin{equation}
    A_{ill} = A_{ill}(max) - A_{ill}(\phi_{T1}, \phi_{T4})
,\end{equation}

\begin{equation}
    A_{ill}(\phi_{T1}, \phi_{T4}) - F_{star} = \delta_{night} + f_{cres}^{transit}
,\end{equation}
and, based on symmetry considerations, the bright crescent at $T_{1}/T_{4}$ will be exactly equal in area to the crescent of the illuminated hemisphere, which is directed away from the observer just before and after occultation, that is, at $t_{1}$/$t_{4}$. In other words: 
\begin{equation}
    f_{cres}^{transit} = f_{cres}^{eclipse}
,\end{equation}

\begin{equation}\label{eqn:deltaNight}
=> \delta_{night} = \delta_{ec} - A_{ill}
.\end{equation}

\subsection{Dayside and nightside temperatures}\label{sec: DayNightTemperatures}
After estimating $\delta_{night}$ with Eq.~\ref{eqn:deltaNight}, nightside temperatures could, in principle, be derived from Eq.~\ref{eqn:ThermalFlux} by replacing $\delta_{therm}$ with $\delta_{night}$ and $T_{d}$ with $T_{n}$. The theoretical dayside and nightside effective temperatures, following  \citet{cowan2011statistics}, are equal to:

\begin{equation}\label{eqn:Dayside}
T_{d} = T_{eff} \sqrt{\frac{R_{\star}}{a}} \left(1 - A_{B}\right)^{\frac{1}{4}} \left(\frac{2}{3} - \frac{5}{12}\epsilon \right)^\frac{1}{4}
,\end{equation}
\begin{equation}\label{eqn:Nightside}
T_{n} = T_{eff} \sqrt{\frac{R_{\star}}{a}} \left(1 - A_{B}\right)^{\frac{1}{4}} \left(\frac{\epsilon}{4} \right)^\frac{1}{4}
,\end{equation}

\noindent
where $A_{B}$ is the Bond albedo and $\epsilon$ is the heat circulation efficiency that parameterizes the heat flow from the dayside to the nightside: $\epsilon = 0$ indicates no heat circulation to the nightside, while $\epsilon = 1$ means perfect heat redistribution. Assuming a zero Bond albedo, that is, a perfectly absorbing exoplanet, the maximum dayside temperature is equal to $T_{d,max}=T_{d~(A_B=0, \epsilon=0)}= T_{eff}~\sqrt{R_{\star}/a}~(2/3)^{1/4}$,  which corresponds to $\epsilon=0$ and thus $T_{n}=0$~K; the lower limit would be the uniform temperature for $\epsilon=1$, namely: $T_{d, uni}=T_{d~(A_B=0, \epsilon=1)}=T_{eff}~\sqrt{R_{\star}/a} ~(1/4)^{1/4}=T_{n}$. 

The maximum value of $A_{g}$ for 100$\%$ reflection and a null thermal emission ($\delta_{ec} = \delta_{ref}$) further permits us to estimate the maximum possible Bond albedo $A_{B}$ by using the relation $A_{B} = 3/2~A_{g}$ for a perfect Lambertian sphere \citep{rowe2006upper}. In this way, the lower limit on $T_{d}$ corresponding to the maximum $A_{B}$ value and $\epsilon=1$ can be computed. However, this relation cannot be assumed when $A_{g}$ itself is close to or greater than 1. Moreover, in the  case of significant nightside emission, we can further constrain the lower limit on $T_{d}$ as $T_{d} \ge T_{n}$, because the dayside cannot be colder than the nightside ($\epsilon$=1). 

 \begin{table*}[h!]
  \centering
   \caption{Best-fit parameters and associated uncertainties from our fit of primary transits, secondary eclipses, and phase variations.}\label{tab:BestfitPars}
    \begin{threeparttable}
   \begin{tabular}{lcccccccccl} \hline\hline
   Planet & $P$ & $T_{c}$ & $T_{dur}$ & $i$ & $R_{p}/R_{\star}$ & $\delta_{ec}$ & $A_{ph}$ \\
       & (days)  & ({$\rm{BJD_{TDB}}$} - 2450000)  & (days) & (degree)&   & (ppm) & (ppm) \\\hline
 \\
   Kepler-10b & 0.83749080(11) & 4964.5751(01) & 0.0751(04) & 83.99 $\pm$ 0.08 & 0.0127(01) & 10.4 $\pm$ 0.9 &  7.4 $\pm$ 0.8  \\\\
   Kepler-78b & 0.35500628(10) & 4953.9599(01) & 0.0343(02) & 79 $\pm$ 1 & 0.0147(10) & 12.4 $\pm$ 1.3 &  $7.6\substack{+1.1\\-0.9}$ \\\\
   Kepler-407b & 0.6693119(55) & 4967.1292(07) & 0.0680(10) & $86.73\substack{+2.09\\-2.24}$ & 0.0103(01) & $ 6.0\substack{+1.9\\-1.6}$ &  $3.2\substack{+1.5\\-1.6}$ \\\\
   K2-141b & 0.2803247(12) & 7744.0716(02) & 0.0401(03) & $86.57\substack{+2.23\\-2.69}$ & 0.0204(01) & $26.2\substack{+3.6\\-3.7}$ & $23.4\substack{+3.9 \\ -4.0}$ \\\\
   K2-131b & 0.3693194(77) & 7627.6213(03) & 0.0428(08) & 82.7$\substack{+4.0\\-3.2}$ & 0.0190$\substack{+0.0004\\-0.0003}$ & 28$\pm$9 & 23$\pm$9 \\\\
   K2-106b & 0.5712919(144) & 7394.0111(11) & 0.0630(12) & $86.34\substack{+2.41\\-2.56}$ & 0.0153(03) & $25.3\substack{+7.7\\-7.6}$ & $16.1 \pm 7.0$ \\\\
   K2-229b & 0.584253(11) & 9784.3558(04) & 0.0536(08)
  & 87.4$\substack{+1.8\\-1.9}$ & 0.0135(02) & 10.6$\substack{+4.2\\-4.1}$ & 4.3$\substack{+4.6\\-3.0}$ \\\\
  K2-312b & 0.719576(18) & 8097.7261(12) & 0.0754(14) & $82.59\substack{+1.18\\-1.11}$ & 0.0121(02) & $8.1\substack{+3.9\\-3.7}$ &  2.7$\pm$3.5 \\\\\hline
  \end{tabular}
  \end{threeparttable}
 \end{table*}

\section{Results }\label{sec: results}

We searched for the secondary eclipse and phase variations for each of the sixteen USP planets that successfully passed our selection criterion described in Sect.~\ref{sec: TargetSelection}. However, we detected the secondary eclipse with $S/N > 2$ only in half of our sample, that is, in the eight systems Kepler-10b, Kepler-78b, Kepler-407b, K2-106b, K2-131b, K2-141b, K2-229b, and K2-312b, which are discussed  sequentially below. 
The secondary eclipse and phase variations went undetected in the other systems mainly because the signal of the optical occultation is actually shallower than our conservative estimates in Sect.~\ref{sec: TargetSelection} and, thus, is buried in the noise. 

\subsection{Kepler-10b}
\begin{figure}[h!]
\centering
\includegraphics[width=\linewidth]{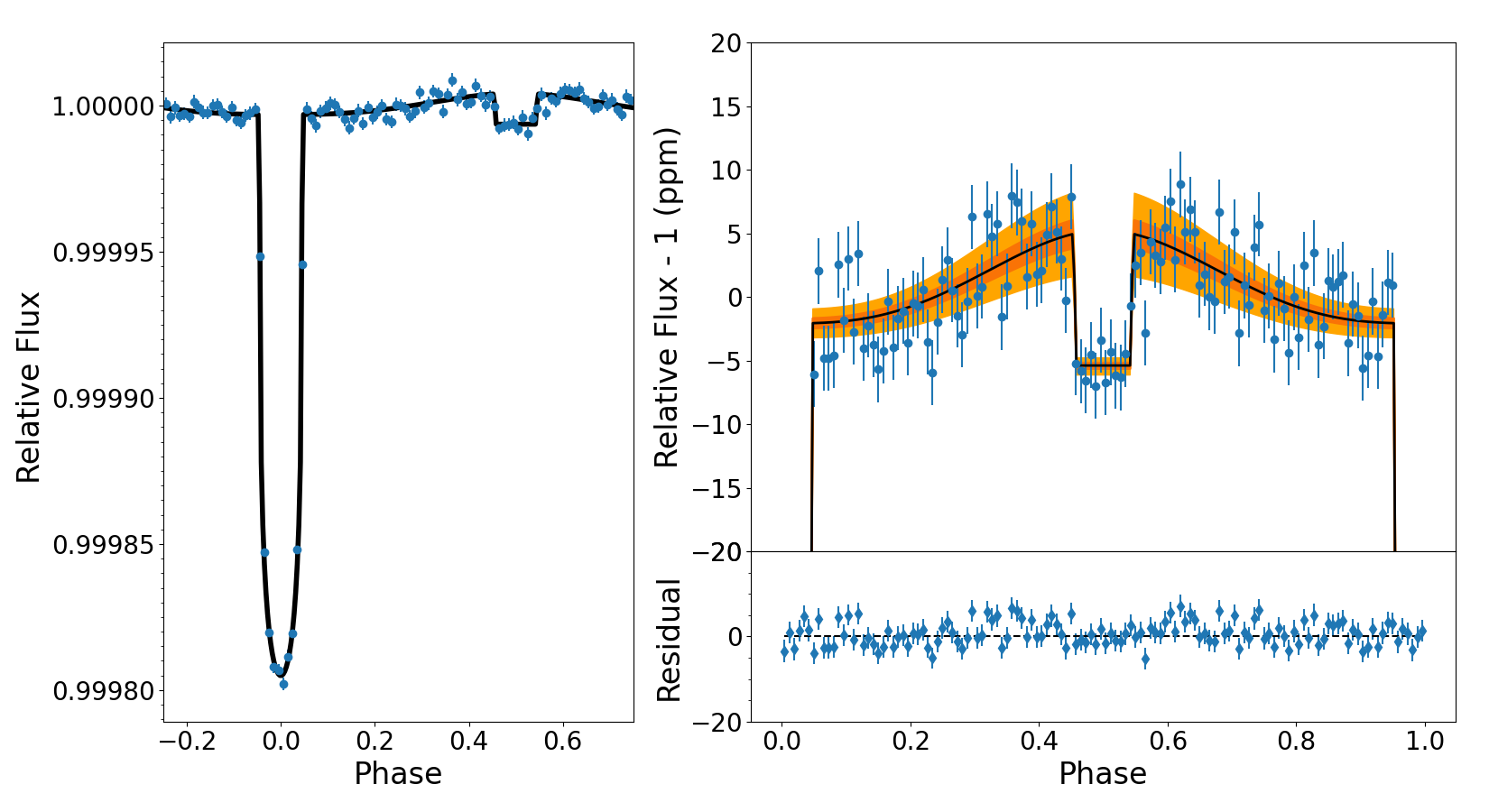}
\hspace{1.0cm}
\caption{Planetary light curve showing the primary transit, secondary eclipse and the phase variation. Left: Detrended, phase-folded, and binned light curve of Kepler-10b (blue points). The best-fit model is over-plotted in solid black line. Right: Enlarged view of the secondary eclipse and phase variation. Uncertainties of the best-fit model are shown at $1\sigma$ and $3\sigma$ in dark and light orange, respectively. The residuals of the best-fit model are displayed in the bottom.}
\label{fig: kepler-10_transit_SE}
\end{figure}

Kepler-10b is the first rocky planet discovered by the Kepler space telescope in 2011 \citep{batalha2011kepler}. It orbits a $\sim$10 Gyr-old solar-like G dwarf with a period of 0.837~d or $\sim$20~hrs, and has a transiting companion with a period of $\sim 45$~d \citep{2011ApJFressin, 2014ApJDumusque, 2016ApJWeiss}. The planet's mass is around 3.6 $M_{\oplus}$ and a radius of 1.48 $R_{\oplus}$. From its bulk density, interior structure models suggest a rocky, Earth-like composition. The best-fit model parameters of our simultaneous fit of transit, secondary eclipse, and phase curve to the \textit{Kepler} short-cadence data are listed in Table~\ref{tab:BestfitPars} and the best-fit model plot is shown in Fig.~\ref{fig: kepler-10_transit_SE}.

The secondary eclipse depth is found to be $10.4\pm 0.9$ ppm and the amplitude of the phase variation $7.4\pm0.8$ ppm. The difference between these two parameters indicates a nightside emission of $3.0\pm1.2$~ppm (Eq. \ref{eqn:deltaNight}), which is significant at the 2.5$\sigma$ level. This fully agrees with the results of \citet{Fogtman-Schultz2014ApJ}, but is slightly at odds with other works reporting negligible nightside emission \citep{Sheets2014ApJ, Hu2015ApJ, Esteves2015ApJ}. The nightside temperature corresponding to our nightside emission is $T_{n}\sim$2800~K. The temperature of the dayside cannot be lower than $T_{n}$ and must be, thus, $T_{d} \ge 2800$~K.

\begin{figure}[h!]
\centering
\includegraphics[width=\linewidth]{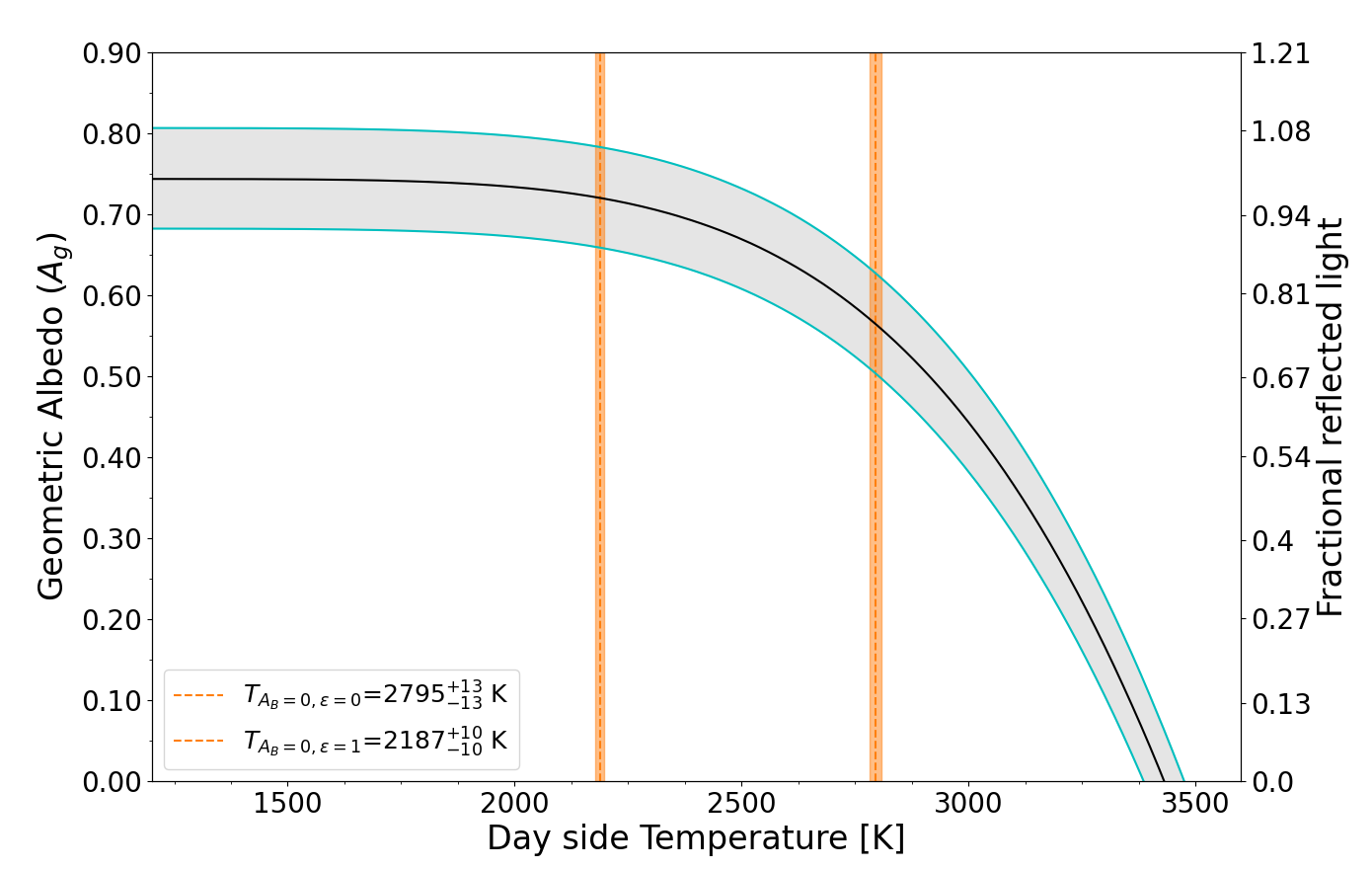}
\hspace{1.0cm}
\caption{Geometric albedo $(A_{g})$ estimated as a function of the dayside temperature of Kepler-10b. The cyan lines correspond to the $3\sigma$ uncertainty in the computation of $A_{g}$. The two vertical red lines and orange regions display the theoretical temperatures and associated $1\sigma$ error bars, by considering zero Bond albedo and negligible or perfect heat redistribution. The maximum geometric albedo is at $0.74\pm0.06$. The maximum dayside temperature can be as high as 3430(+50/-40)~K in the case of 100$\%$ thermal emission.}
\label{fig: kepler-10b_GA}
\end{figure}

From $\delta_{ec}$, we computed the geometric albedo as a function of the dayside temperature (using Eq. \ref{eqn:Albedo} and \ref{eqn:ThermalFlux}). The $A_{g}$ vs $T_{d}$ plot is shown in Fig.~\ref{fig: kepler-10b_GA}. The maximum $A_{g}$ value corresponds to a relatively high value of $0.74\pm0.06$, but $A_{g}$ should be lower than $\sim0.6$ from the previous estimate on the dayside temperature of $T_{d} \ge 2800$~K (Fig.~\ref{fig: kepler-10b_GA}). On the contrary, the maximum dayside temperature for a perfectly absorbing planet ($A_{g}=0$) is $3430\substack{+50\\-40}$~K (see Fig.~\ref{fig: kepler-10b_GA} for zero fractional reflected light). The two theoretical dayside temperatures, namely $T_{d, max}=T_{d~(A_B=0, \epsilon=1)}$ and $T_{d, uni}=T_{d~(A_B=0, \epsilon=1)}$ (see Sect.~\ref{sec: DayNightTemperatures}), are also shown in the plot.

\subsection{Kepler-78b}
\begin{figure}[h!]
\centering
\includegraphics[width=\linewidth]{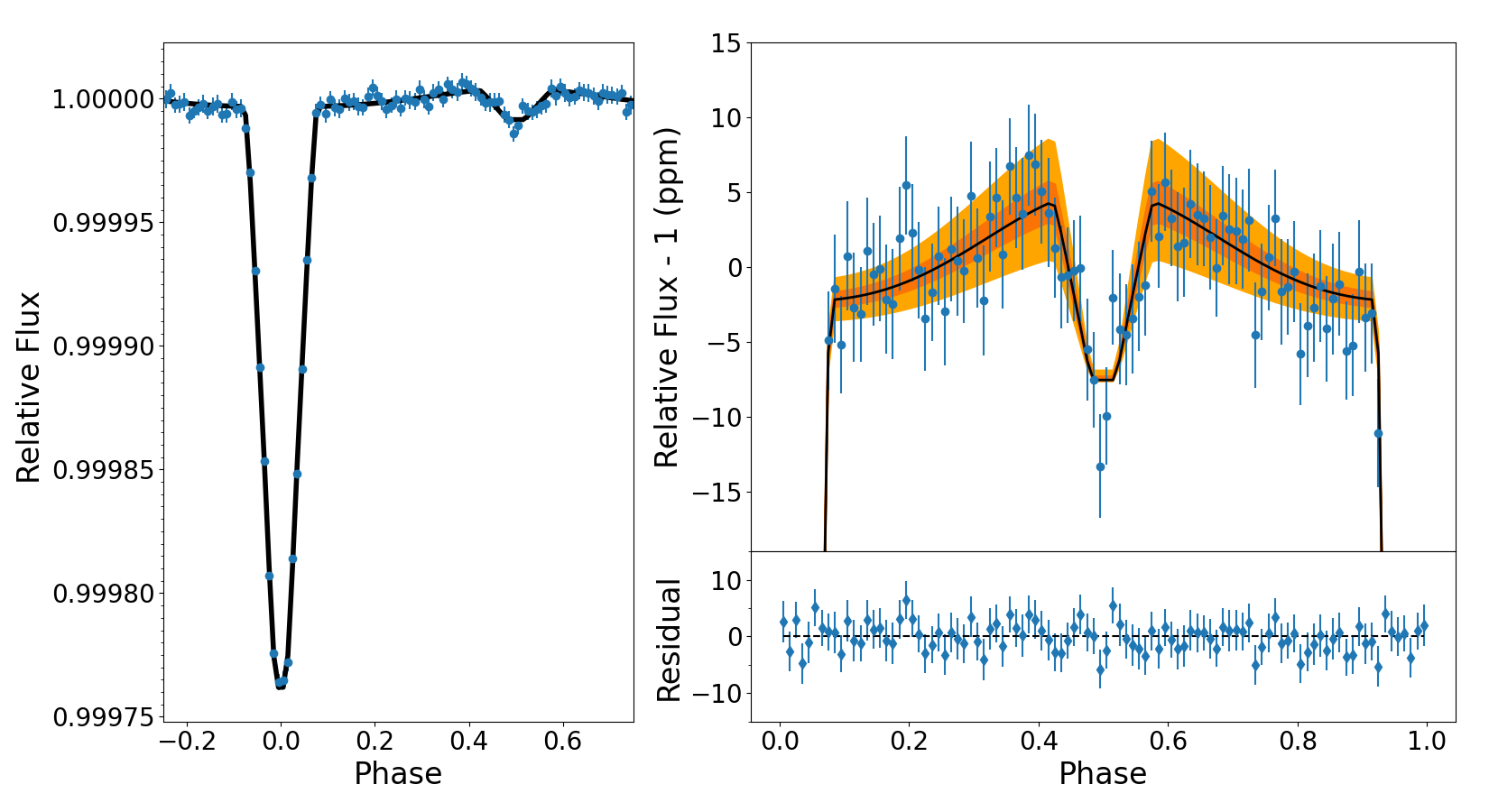}
\hspace{1.0cm}
\caption{Primary transit, secondary eclipse, and the phase variation of Kepler-78b (ref. Fig.~\ref{fig: kepler-10_transit_SE}).}
\label{fig: kepler-78_transit_SE}
\end{figure}

Kepler-78b is a rocky super-Earth orbiting a relatively young G-type dwarf with an age of $\sim$750~Myr old in an 8.5~hr orbit \citep{sanchis2013transits, Howard_2013Nature, Pepe2013}. The planet has a mass of 1.8 $M_{\oplus}$ and a radius of 1.2 $R_{\oplus}$ resulting in an Earth-like density of 5.3 g/cc. Short-cadence data for this target is not available and therefore we used the four-year long-cadence light curve for our analysis. Our best-fit model is shown in Fig.~\ref{fig: kepler-78_transit_SE}. 

\begin{figure}[h!]
\centering
\includegraphics[width=\linewidth]{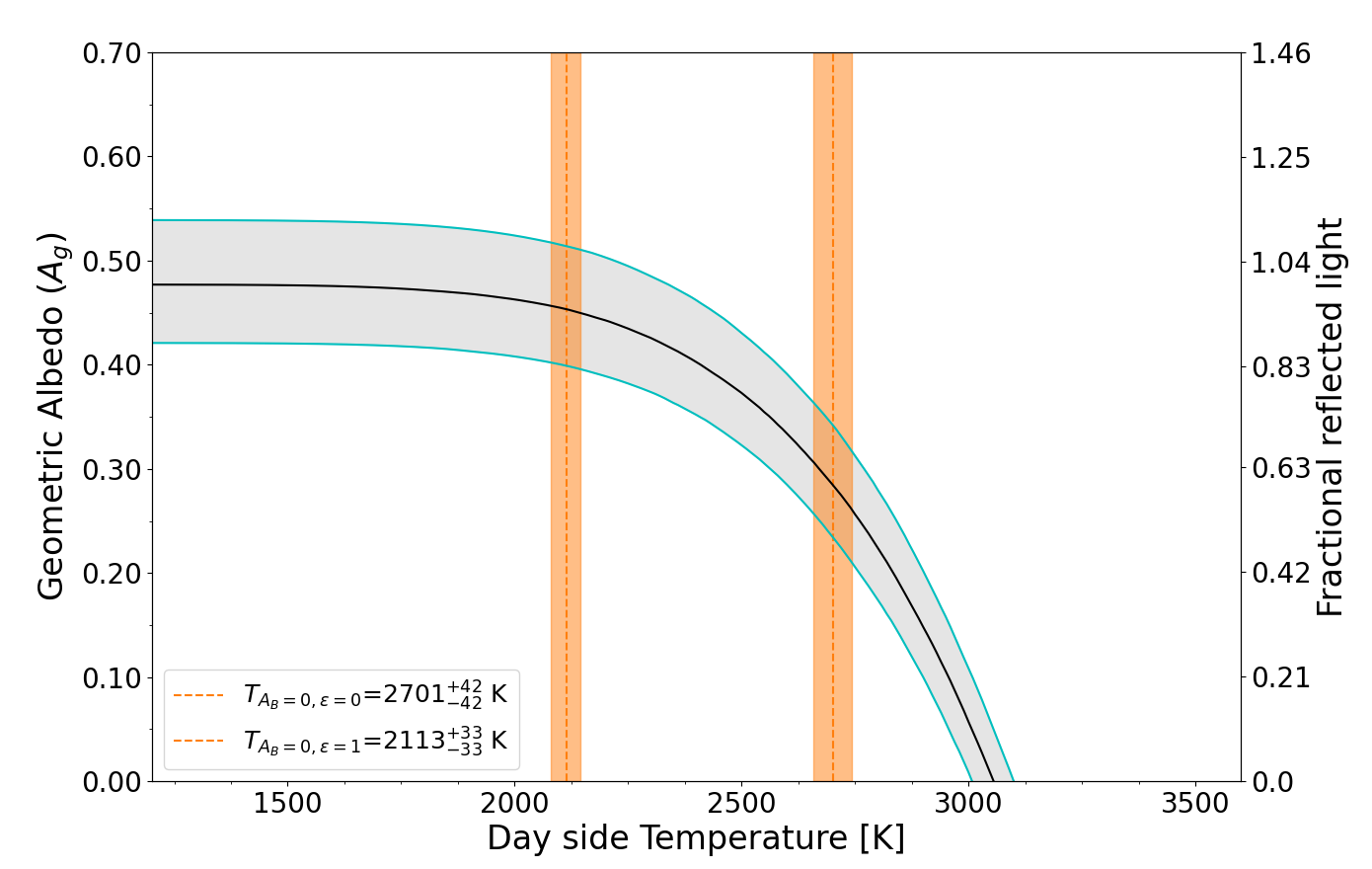}
\hspace{1.0cm}
\caption{Constraints on the geometric albedo and the dayside temperature of Kepler-78b (ref. Fig.~\ref{fig: kepler-10b_GA}). The maximum geometric albedo is 0.48$\pm$0.06. The maximum dayside temperature is 3060(+40/-50)~K in the case of $100\%$ thermal emission.}
\label{fig: kepler-78b_GA}
\end{figure}

Our analysis shows a high-confidence secondary eclipse detection with $\delta_{ec}=12.4\pm1.3$~ppm and a phase-amplitude of $7.6\pm 1.0$~ppm and, hence, similar to Kepler-10b, the difference implies a nightside emission of $\delta_{night}=4.8 \pm 1.6$~ppm ($3\sigma$). Compared to the results of \citet{sanchis2013transits}, that is, $\delta_{ec}=10.5\pm1.2$~ppm and $A_{ill}=8.8\pm1.0$~ppm,
we obtained a slightly deeper secondary eclipse and a slightly lower amplitude of phase variation, although our solution and theirs agree within $2\sigma$. The nightside emission corresponds to a temperature $T_{n}$ $\sim2700$~K and, hence, it gives a lower limit to the dayside temperature, $T_{d} \gtrsim 2700$~K). The relation between $A_{g}$ and $T_{d}$ for the measured $\delta_{ec}$ is shown in Fig.~\ref{fig: kepler-78b_GA}; $A_{g}$ tends to $0.48\pm0.06$ at the saturation level, that is, for the lower dayside temperatures, while the maximum dayside temperature for a null reflection ($A_{g} = 0$) is 3060$\substack{+40\\-50}$ K. Using the lower limit on the dayside temperature, we further constrain the geometric albedo to be less than 0.3. 

\subsection{Kepler-407b}
\begin{figure}[h!]
\centering
\includegraphics[width=\linewidth]{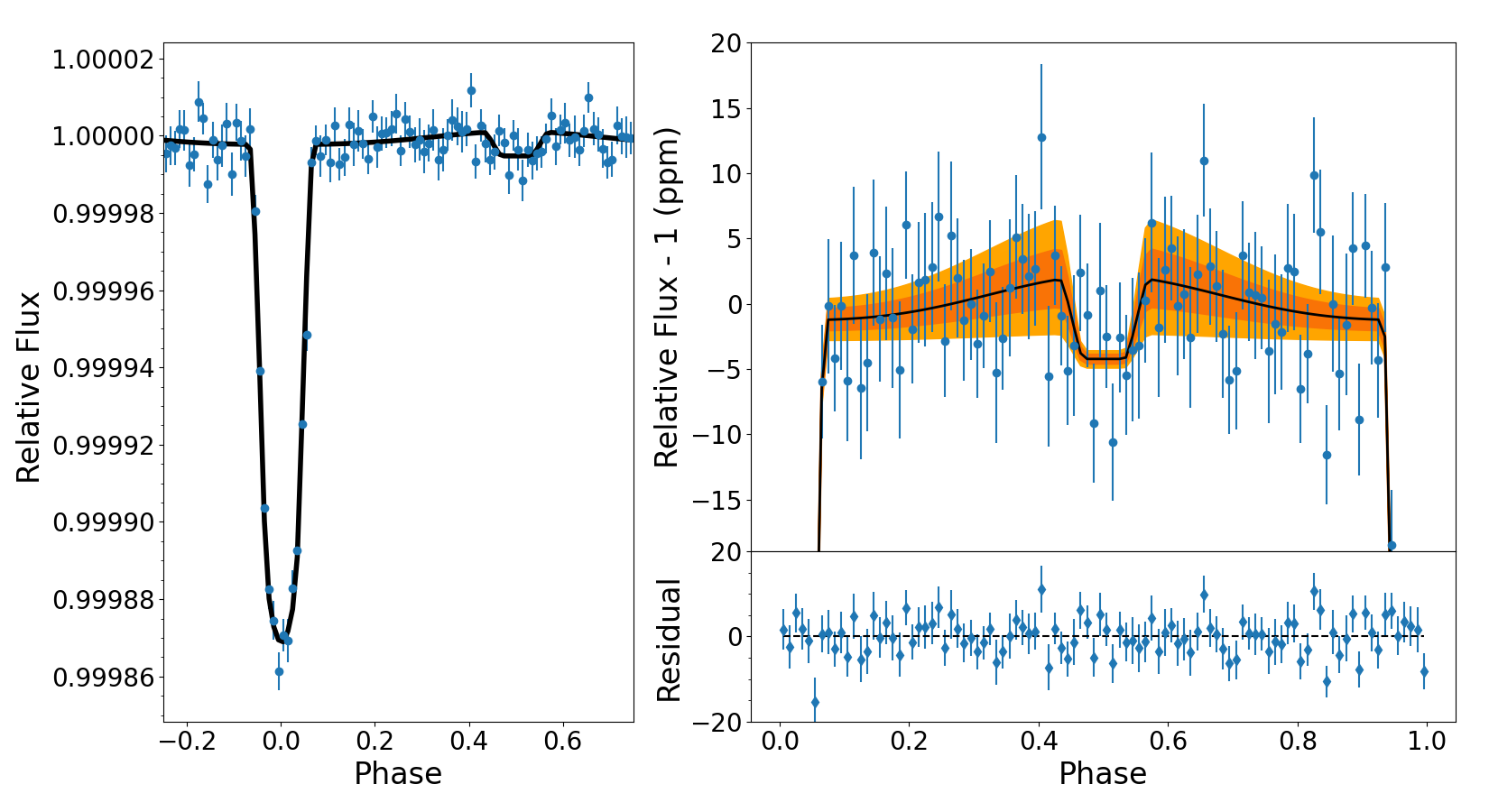}
\hspace{1.0cm}
\caption{Primary transit, secondary eclipse, and the phase variation of Kepler-407b (ref. Fig.~\ref{fig: kepler-10_transit_SE}).} 
\label{fig: kepler-407_transit_SE}
\end{figure}

Kepler-407b orbits a G dwarf in a 16~hr period. Previous RV analysis on this target provided an upper limit on the planet's mass and also a partial orbit of a non-transiting companion \citep{Marcy2014ApJ}. We  determined its radius precisely at $1.07\pm0.02$ $R_{\oplus}$. Furthermore, we present a $3\sigma$ detection of a secondary eclipse depth at $6\pm2$ ppm and a phase variation with an amplitude at $3.2\pm1.5$ ppm. The difference of these two parameters is positive, but not precise enough to claim evidence for a significant nightside emission. The best-fit model plot is shown in Fig.~\ref{fig: kepler-407_transit_SE}.    

\begin{figure}[h!]
\centering
\includegraphics[width=\linewidth]{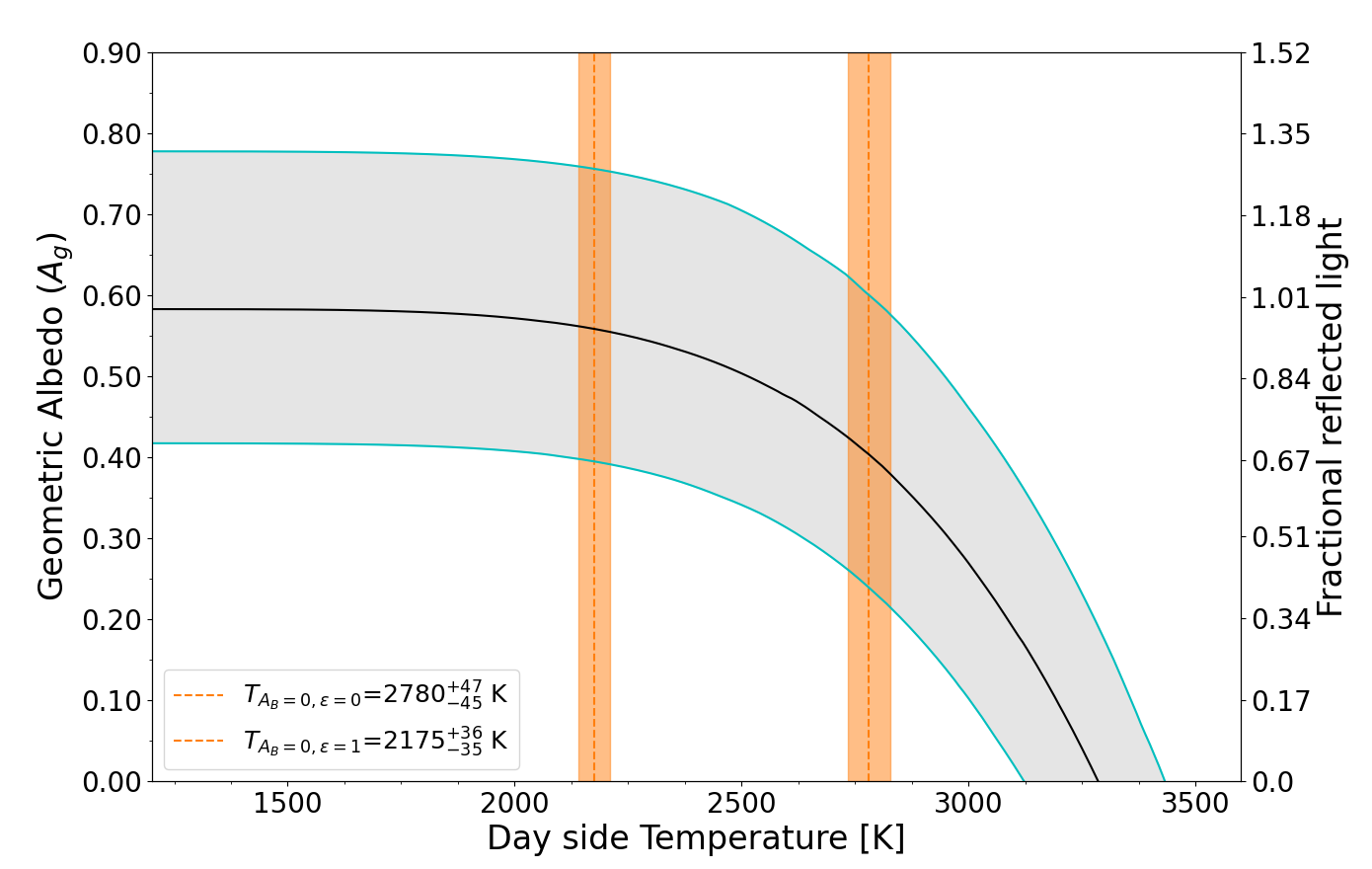}
\hspace{1.0cm}
\caption{Constraints on the geometric albedo and the dayside temperature of Kepler-407b (ref. Fig.~\ref{fig: kepler-10b_GA}). The geometric albedo at saturation is 0.56(+0.19/-0.18), while the maximum dayside temperature is 3270$\pm$140~K.}
\label{fig: kepler-407_GA}
\end{figure}

Using the secondary eclipse depth value, we computed $A_{g}$ as a function of $T_{d}$ (see Fig.~\ref{fig: kepler-407_GA}). The maximum geometric albedo, in the case where the eclipse depth is due to the pure reflection of the stellar light, is $0.56\pm0.19$. By assuming isotropic scattering from the planet, that is, $A_{B} = 3/2 A_{g}$, we obtained lower limits on the dayside temperature as $T_{d} \ge 1400$~K and $T_{d} \ge 1800$~K for $\epsilon=1$ and $\epsilon=0$, respectively. In the case of pure thermal emission, the dayside temperature could be as high as 3270 $\pm$ 140 K. 

\subsection{K2-141b}
\begin{figure}[h!]
\centering
\includegraphics[width=\linewidth]{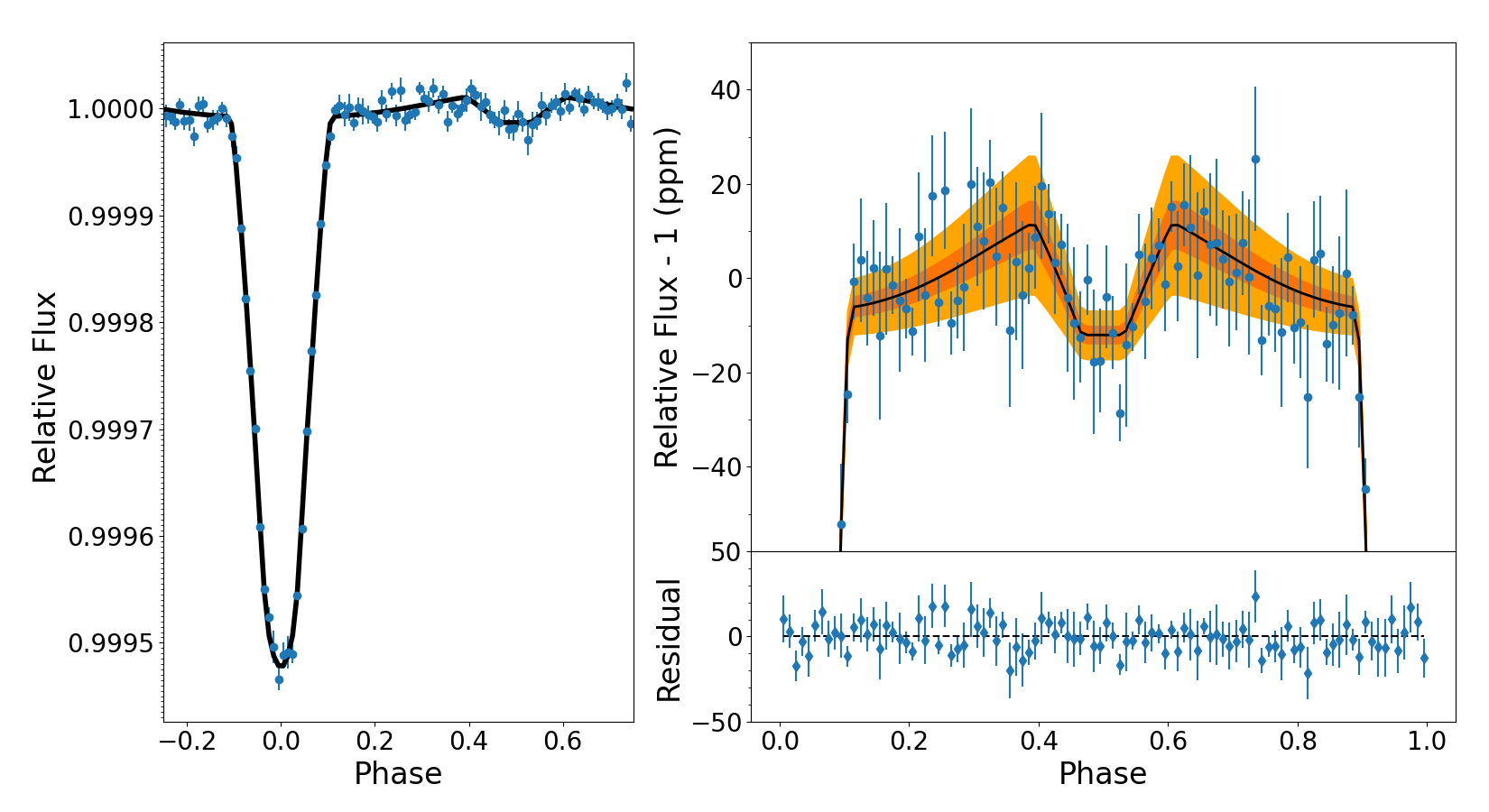}
\hspace{1.0cm}
\caption{Primary transit, secondary eclipse, and the phase variation of K2-141b (ref. Fig.~\ref{fig: kepler-10_transit_SE}).}
\label{fig: k2-141b}
\end{figure}

K2-141b is a rocky super-Earth in a 6.7~hr orbit around an active K-dwarf with a rotation period of $\sim$14~d \citep{malavolta2018ultra, Barragan2018A&A}. The planet has a mass of 5.1 $M_{\oplus}$ \citep{malavolta2018ultra} and a radius of 1.5 $R_{\oplus}$, thereby resulting in a super-terrestrial density of 8.2 g/cc. Our analysis provides a robust detection ($>6\sigma$) of both the eclipse depth at 26.2$\substack{+3.6\\-3.8}$~ppm and the phase variation with an amplitude of 23.4$\substack{+3.9\\-3.9}$~ppm, in agreement with \citet{malavolta2018ultra}. Since $\delta_{ec}$ and $A_{ill}$ are indistinguishable within the error bars, there is no indication of thermal emission from the nightside. Figure~\ref{fig: k2-141b} shows the phase-folded transit, eclipse, and phase variation along with the best-fit model. 

\begin{figure}[h!]
\centering
\includegraphics[width=\linewidth]{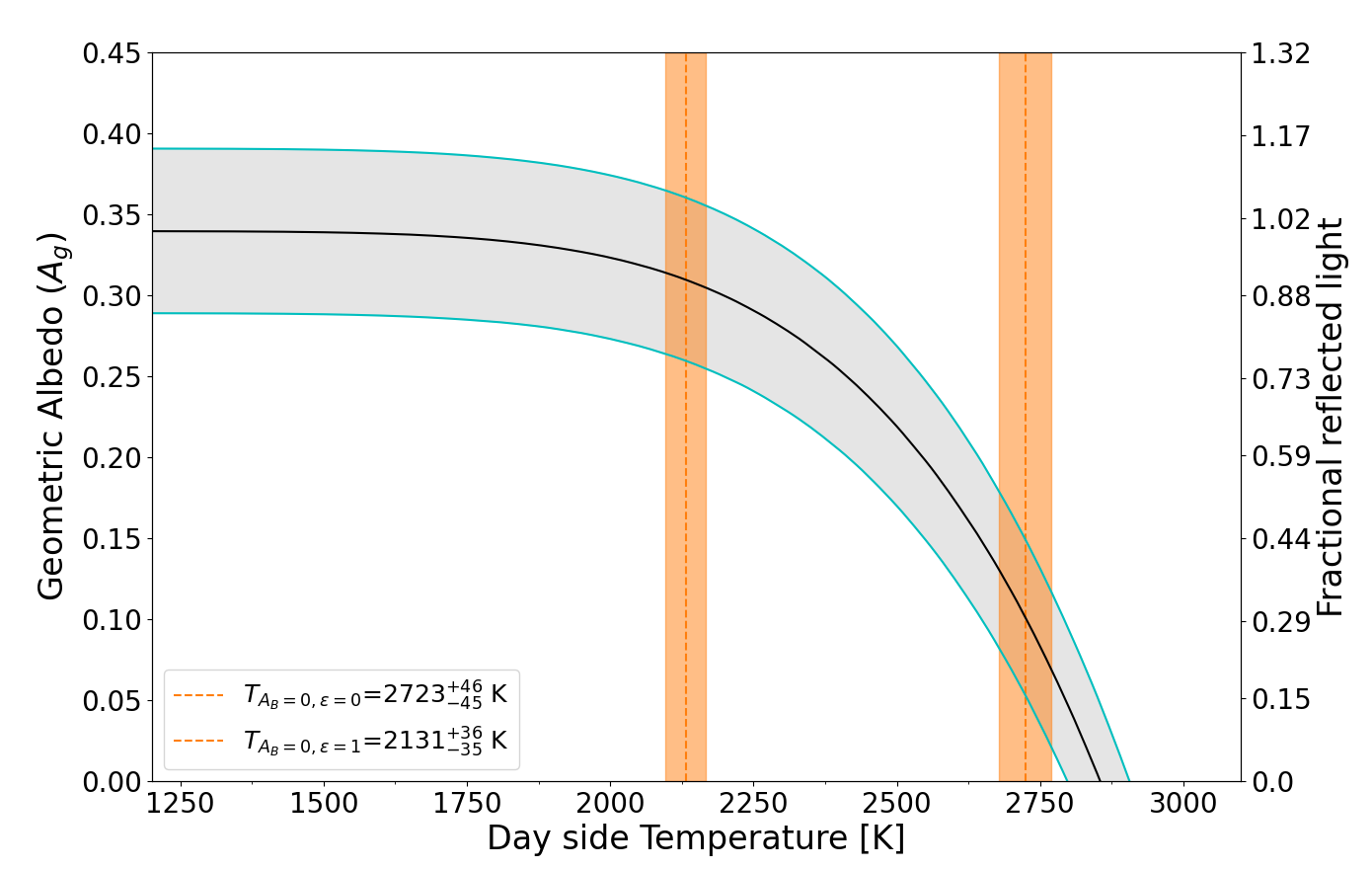}
\hspace{1.0cm}
\caption{Constraints on the geometric albedo and the dayside temperature of K2-141b (ref. Fig.~\ref{fig: kepler-10b_GA}).  The maximum geometric albedo is $0.34\pm0.05$, while the maximum dayside temperature is 2860(+50/-60)~K.}
\label{fig: k2-141b_GA}
\end{figure}

Given the observed eclipse depth ($\delta_{ec}$), we computed $A_{g}$ as a function of $T_{d}$ (Fig.~\ref{fig: k2-141b_GA}). The geometric albedo saturates at $0.34\pm0.05$. Assuming isotropic scattering for $A_{g}=0.34$, we obtain an upper limit of 0.5 on the Bond albedo and, therefore, the theoretical lower limit on the dayside temperature is computed to be $\sim2300$~K for $\epsilon=0$. The maximum dayside temperature that the planet could achieve for no reflection is $2860\substack{+50\\-60}$~K. 
 
\subsection{K2-131b}
\begin{figure}[h!]
\centering
\includegraphics[width=\linewidth]{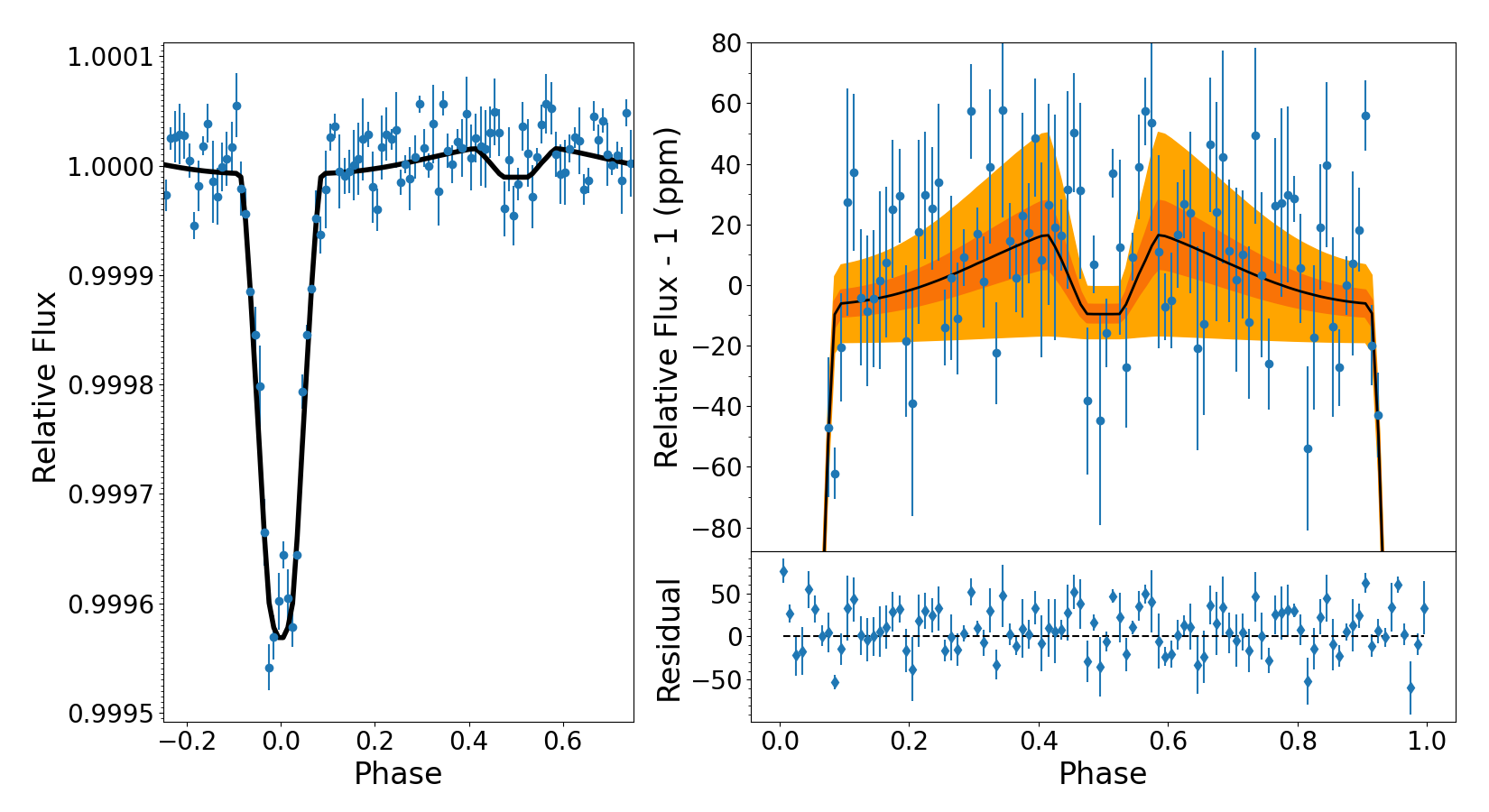}
\hspace{1.0cm}
\caption{Primary transit, secondary eclipse, and the phase variation of K2-131b (ref. Fig.~\ref{fig: kepler-10_transit_SE}).}
\label{fig: k2-131b}
\end{figure}

The K2-131 system consists of a single discovered exoplanet in an 8.86~hr orbit around an active late G main-sequence star with a rotation period of $\sim$11 days \citep{2017AJDai}. The planet's mass of 6.3 $M_{\oplus}$ \citep{Dai2019} and a radius of 1.6 $R_{\oplus}$ results in a similar composition as that of K2-141b. The simultaneous model of transit, eclipse, and phase variation corresponding to the best-fit parameters is shown in Fig.~\ref{fig: k2-131b}.

With a confidence level slightly higher than $3\sigma$, we found a secondary eclipse depth of 28.3$\substack{+8.9\\-9.0}$ ppm and an amplitude of the phase variation of 27.4$\substack{+8.5\\-8.2}$ ppm. These two parameters are practically equal, which implies that the nightside emission is likely negligible and thus an inefficient heat redistribution.

\begin{figure}[h!]
\centering
\includegraphics[width=\linewidth]{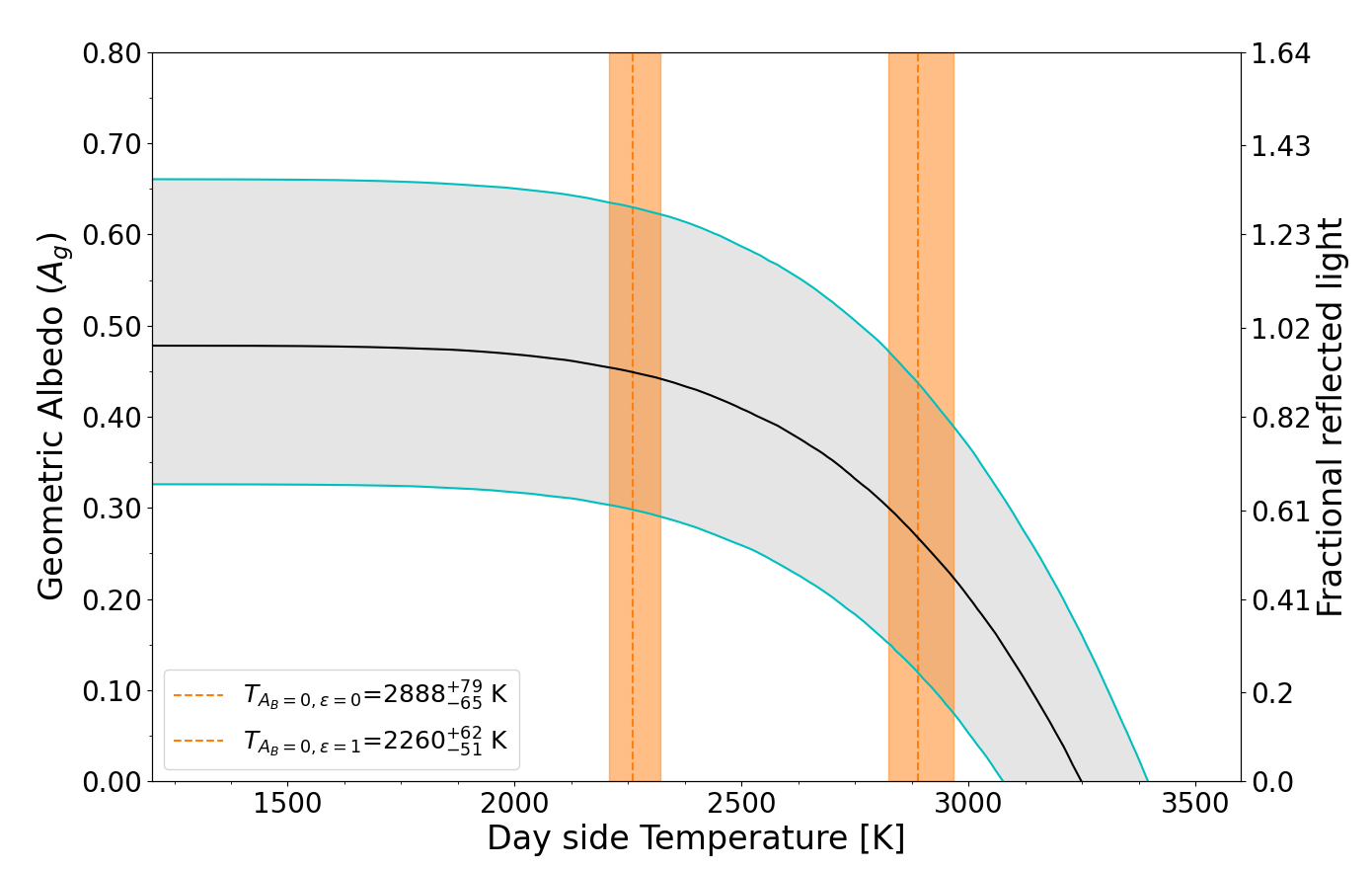}
\hspace{1.0cm}
\caption{Constraints on the geometric albedo and the dayside temperature of K2-131b (ref. Fig.~\ref{fig: kepler-10b_GA}). The maximum geometric albedo is 0.55(+0.18/-0.17), while the maximum dayside temperature is 3320(+140/-180)~K.}
\label{fig: k2-131b_GA}
\end{figure}

The variation of $A_{g}$ with $T_{d}$ in Fig.~\ref{fig: k2-131b_GA} shows that the maximum attainable dayside temperature is $3320\substack{+140\\-180}$~K. The geometric albedo at saturation is $0.55\substack{+0.20\\-0.18}$ and the corresponding lower limit on $T_{d}$ for isotropic scattering and $\epsilon=0$ is $\sim$1900~K. 

\subsection{K2-106 b}
\begin{figure}[h!]
\centering
\includegraphics[width=\linewidth]{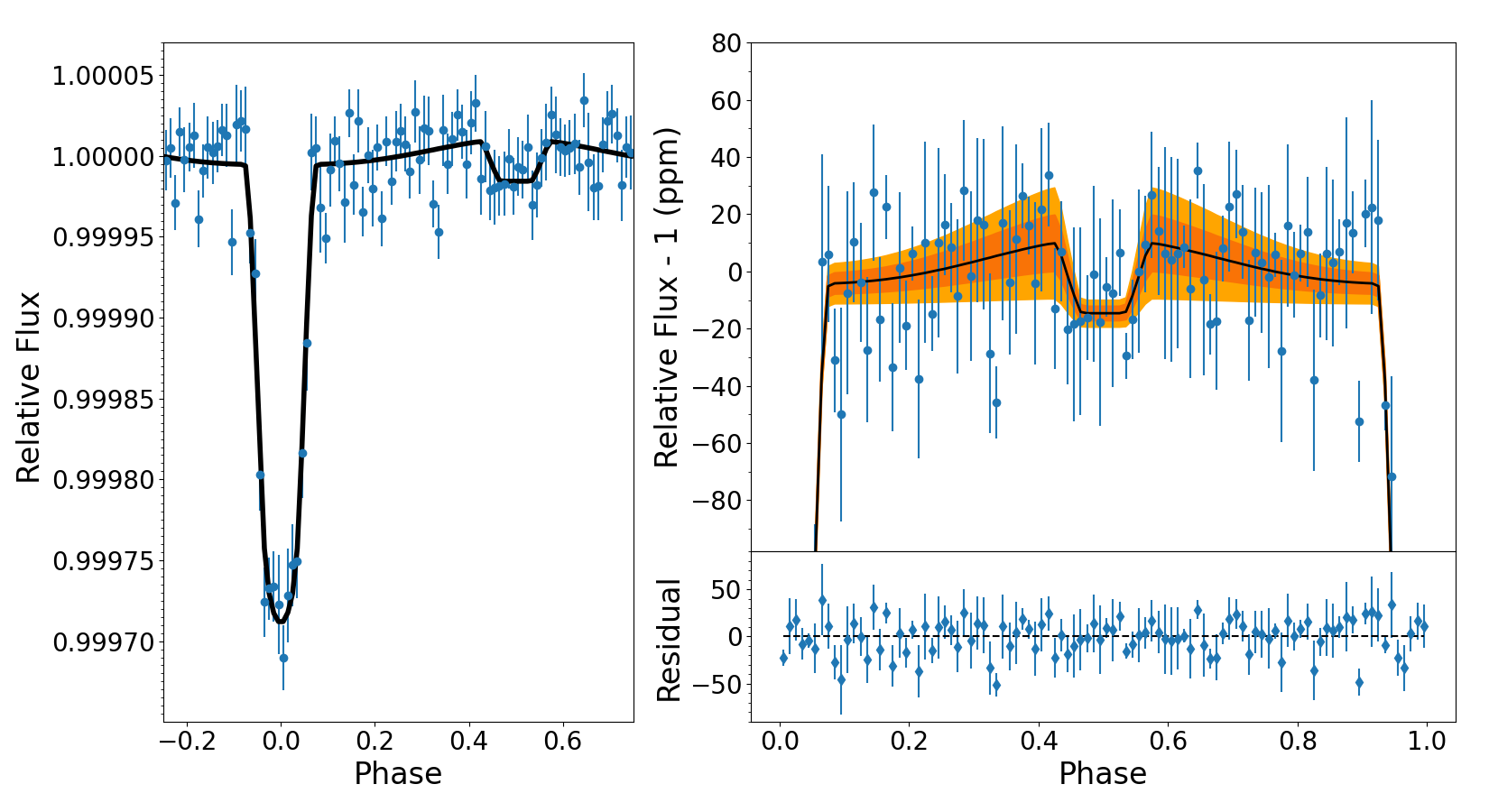}
\hspace{1.0cm}
\caption{Primary transit, secondary eclipse, and the phase variation of K2-106b with $1\sigma$  and $2\sigma$ model uncertainties (ref. Fig.~\ref{fig: kepler-10_transit_SE}).}
\label{fig: k2-106b}
\end{figure}

K2-106b orbits a G dwarf with $T_{eff}\sim$5600~K in a 13.7~hr orbit and is accompanied by a warm Neptune with a period of 13.34~d \citep{adams2017AJ}. With a mass of $8.4~M_{\oplus}  $\citep{2017AJSinuKoff} and a radius of $1.71~R_{\oplus}$, K2-106b is characterized as a dense rocky planet. Our best-fit model for the secondary eclipse and phase variations is shown in Fig.~\ref{fig: k2-106b}. 

\begin{figure}[h!]
\centering
\includegraphics[width=\linewidth]{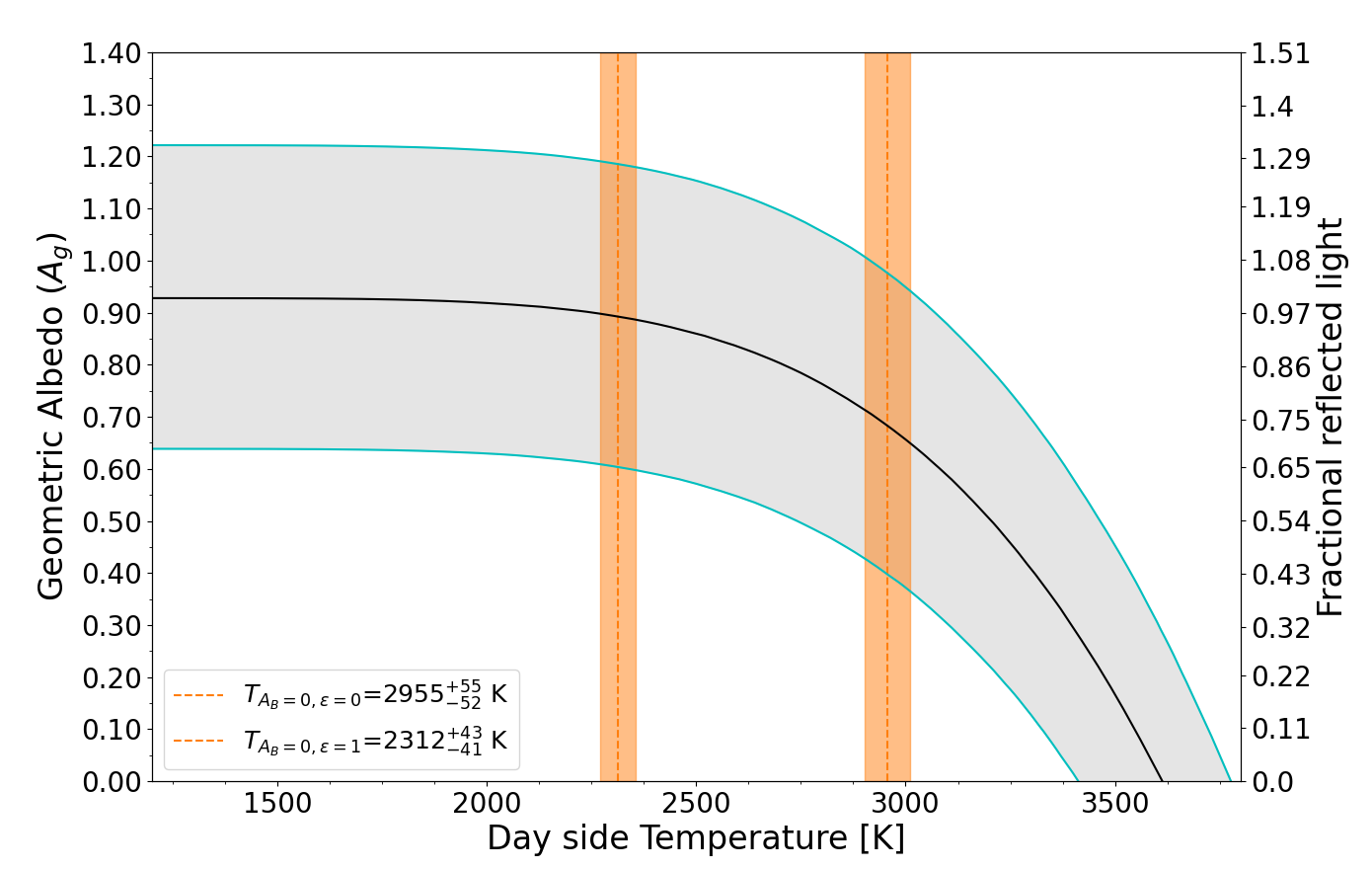}
\hspace{1.0cm}
\caption{Constraints on the geometric albedo and the dayside temperature of K2-106b (ref. Fig.~\ref{fig: kepler-10b_GA}). The maximum geometric albedo is $0.9\pm0.3$, while the maximum dayside temperature is 3620(+160/-200)~K.}
\label{fig: k2-106b_GA}
\end{figure}

We observed a secondary eclipse depth of $25.3\substack{+7.7\\-7.6}$ ppm ($3.3\sigma$) and a phase variation amplitude of $16.1\pm7.0$ ppm ($2.3\sigma$). The positive difference between the two parameters, namely, $9.3\substack{+7.4\\-7.3}$~ppm, may suggest the presence of nightside emission, but the large uncertainty prevents us from drawing any firm conclusion. The $A_{g}$-$T_{d}$ plot shown in Fig.~\ref{fig: k2-106b_GA} suggests a relatively high maximum geometric albedo of 0.9$\pm$0.3. On the contrary, for $100\%$ thermal emission, the maximum dayside temperature is 3620$\substack{+160\\-200}$~K, which is higher than the theoretical maximum value $2955\substack{+56\\-53}$~K for zero Bond albedo.

\subsection{K2-229 b}
\begin{figure}[h!]
\centering
\includegraphics[width=\linewidth]{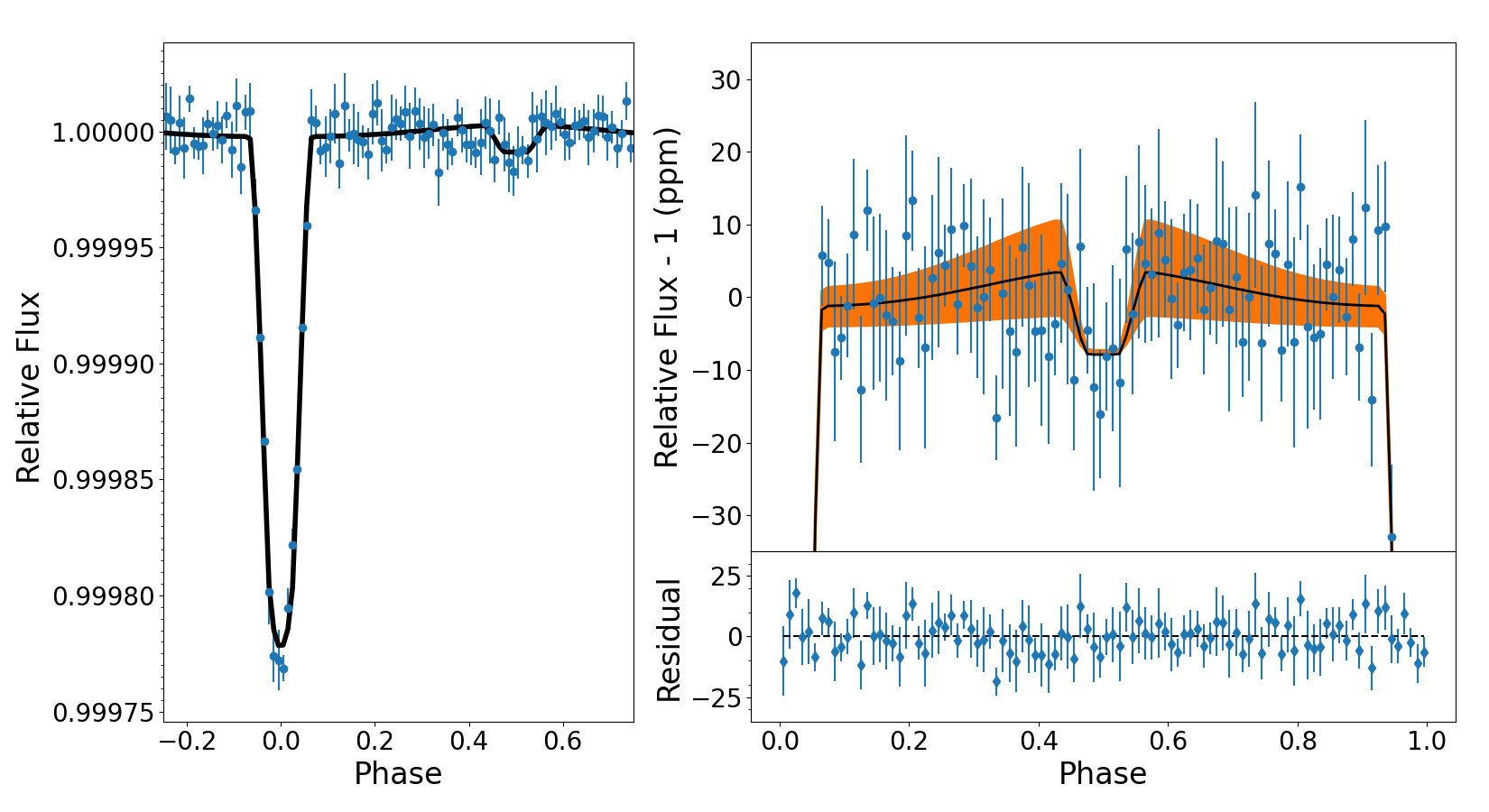}
\hspace{1.0cm}
\caption{Primary transit, secondary eclipse, and the phase variation of K2-229b with $1\sigma$ model uncertainties (ref. Fig.~\ref{fig: kepler-10_transit_SE}).}
\label{fig: k2-229b}
\end{figure}

K2-229b orbits an active late G dwarf in a $\sim$14~hr orbit. With a mass of 2.59 $M_{\oplus}$ and a radius of 1.14 $R_{\oplus}$, it is believed to have a Mercury-like composition \citep{Santerne2018Nat}. Our simultaneous analysis provides a secondary eclipse depth $\delta_{ec}=10.6\substack{+4.2\\-4.1}$~ppm at more than $2\sigma$ confidence level with an amplitude of phase variation being consistent with zero $A_{ill}=4.2\substack{+4.7\\-3.0}$~ppm. Not much can be inferred about the nightside emission due to the poor precision on both $\delta_{ec}$ and $A_{ill}$. The phase-folded light curve, along with the best-fit model, is shown in Fig.~\ref{fig: k2-229b}.

\begin{figure}[h!]
\centering
\includegraphics[width=\linewidth]{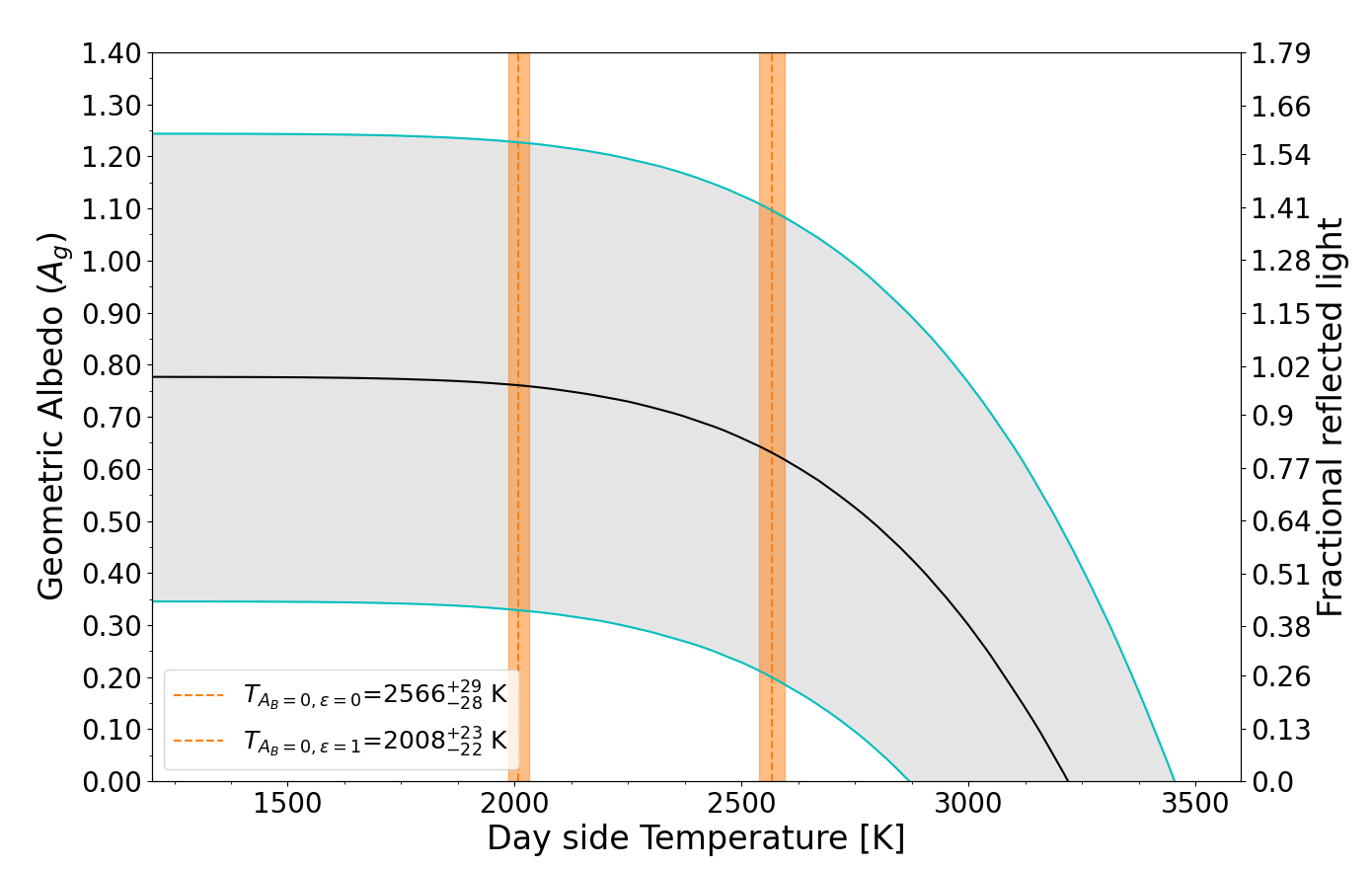}
\hspace{1.0cm}
\caption{Constraints on the geometric albedo and the dayside temperature of K2-229b (ref. Fig.~\ref{fig: kepler-10b_GA}). The geometric albedo at saturation is $0.7\pm0.3$, while the maximum dayside temperature is 3200~K.}
\label{fig: k2-229b_GA}
\end{figure}

The $A_{g}$-$T_{d}$ diagram in Fig. \ref{fig: k2-229b_GA} indicates a maximum geometric albedo of $0.72\substack{+0.33\\-0.31}$, while the maximum dayside temperature is $3200\substack{+180\\-240}$~K. These outcomes should be taken with caution, because the model selection performed in Sect.~\ref{sec: Model_selection} does not favor the secondary eclipse model for K2-229b. More analyses will thus be needed to properly assess the robustness of the K2-229b occultation signal.
 
\subsection{K2-312 b}
\begin{figure}[h!]
\centering
\includegraphics[width=\linewidth]{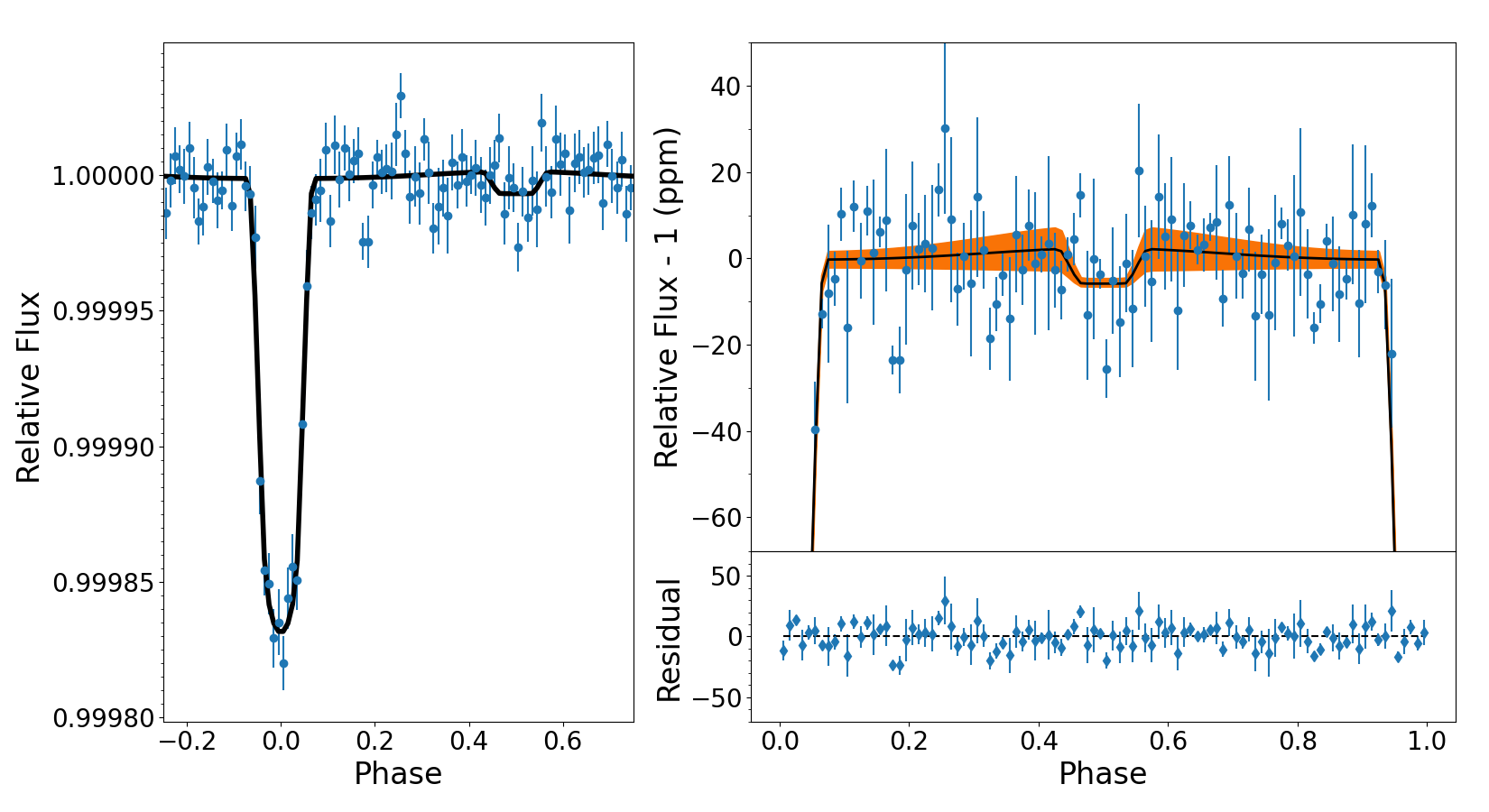}
\hspace{1.0cm}
\caption{Primary transit, secondary eclipse and the phase variation of K2-312b with $1\sigma$ model uncertainties (ref. Fig.~\ref{fig: kepler-10_transit_SE}).}
\label{fig: k2-312b}
\end{figure}

K2-312b orbits an active F main-sequence star in a $\sim$17~hr orbit. The planet's mass and radius is 5.6 $M_{\oplus}$ and 1.61 $R_{\oplus}$, respectively, thereby suggesting a rocky Earth-like composition with no thick atmosphere \citep{2020A&AFrustagli}. Our analysis indicates a secondary eclipse depth of $\delta_{ec} = 8.1\pm3.7$~ppm at a confidence level of 2.2~$\sigma$. However, we could not detect the phase curve variation as its amplitude is consistent with zero ($2.7 \pm 3.5$ ppm). The plot of the best-fit transit, secondary eclipse, and phase curve model over the phase-folded binned data is shown in Fig.~\ref{fig: k2-312b}.

\begin{figure}[h!]
\centering
\includegraphics[width=\linewidth]{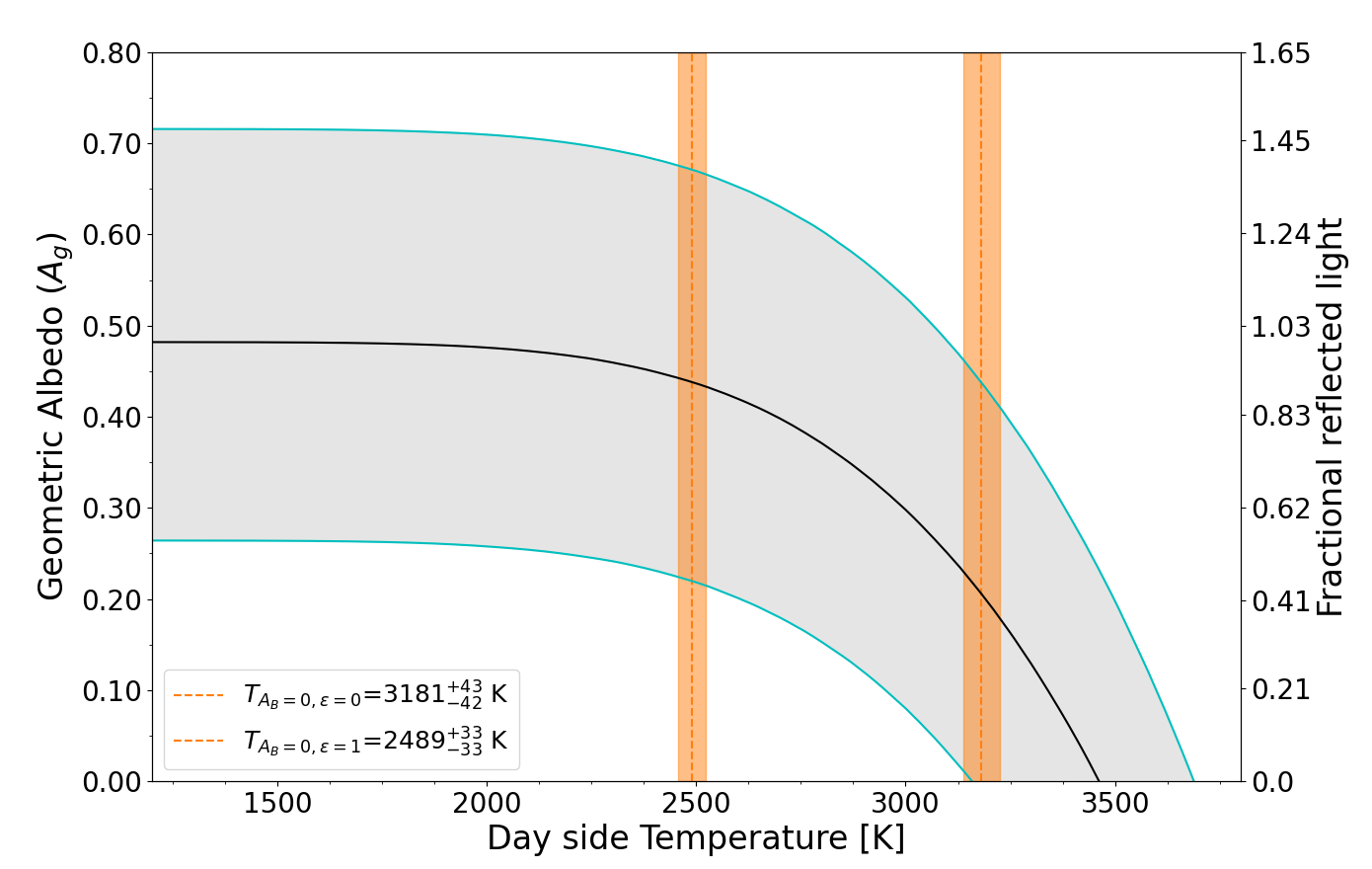}
\hspace{1.0cm}
\caption{Constraints on the geometric albedo, and the dayside temperature of K2-312b (ref. Fig.~\ref{fig: kepler-10b_GA}). The maximum geometric albedo is $0.5\pm0.2$, while the maximum dayside temperature is 3476(+228/-305)~K.}
\label{fig: k2-312b_GA}
\end{figure}

Despite the large uncertainty on $\delta_{ec}$, we estimated $A_{g}$ as a function of $T_{d}$. Figure~\ref{fig: k2-312b_GA} shows that the maximum dayside temperature is 3476$\substack{+228\\-305}$~K for 100$\%$ thermal emission. On the other end, the upper limit on the geometric albedo for 100\% reflection is $0.5\pm0.2$.

\subsection{Summary and model selection using the 
Akaike Information Criterion}\label{sec: Model_selection}
By using the publicly available high-precision \textit{Kepler} photometry, we modeled the secondary eclipses and phase curve variations of eight ultra-short-period sub-Neptunes. We confirm previous detections of secondary eclipse for the three USP planets Kepler-10b, Kepler-78b, and K2-141b, and a marginal detection for K2-312b. We report four new discoveries of secondary eclipses for the planets Kepler-407b ($3.0\sigma$), K2-106b ($3.3\sigma$), K2-131b ($3.2\sigma$), and hints toward K2-229b ($2.5\sigma$) and K2-312b($2.2\sigma$), however, with relatively low significance given the very shallow signals. We also detected the phase variations with confidence levels from 2 to $10\sigma$ for all planets, except K2-229b and K2-312b.

For the low-significance signals, namely, all the systems but Kepler-10, Kepler-78, and K2-141, we computed the values of the Akaike Information Criterion (AIC) for i) the model with the secondary eclipse and phase variation, except for K2-312 and K2-229, for which we only considered the eclipse depth model, as the amplitude of their phase variation is not significant and ii) a constant (flat) model. The $\Delta \rm{AIC}$ values for Kepler-407b, K2-131b, K2-106b, and K2-312b in favor of the planetary model are equal to 6.4, 9.4, 10.0, 9.2, respectively, and would thus indicate "strong evidence” for the presence of the secondary eclipse and phase variation  \citep{kass1995bayes}. None of these $\Delta \rm{AIC}$ values is so high to claim “very strong evidence” ($\Delta \rm{AIC}>10$), which is, however, expected for secondary eclipses detected at the $\sim2-3\sigma$ level. In constrast, however, the $\Delta \rm{AIC}\sim1$ for K2-229b does not provide any “positive evidence” in favor of the secondary eclipse model. We interpret this as a warning that the detection of the secondary eclipse of K2-229b may be spurious. Further analyses, going beyond the scope of this paper, are needed to investigate the occultation signal of this planet.

The AIC model selection is just as much a Bayesian procedure as is the BIC model selection \citep{burnham2004multimodel}. However, in our specific cases, the BIC may penalize complex models (e.g., phase curve and secondary eclipse models against flat out of transit models) more strongly than the AIC. This is due to the large number of data points in the light curves and the very small planetary signals with amplitudes less than the scatter of the data.

From the measured occultation depths, we estimated the geometric albedo $A_{g}$ as a function of the average dayside temperature, $T_{d}$, for each planet. We then provided an upper limit on both $A_{g}$ and $T_{d}$ in the case of a purely reflective ($100\%$ reflection) or purely absorbing ($100\%$ thermal emission) planet, respectively.

\section{Discussion and conclusion}\label{sec: Discussions}

\noindent
Optical data only cannot break the degeneracy between reflected and thermally emitted light. Nevertheless, the analysis of the optical \emph{Kepler} photometry in the present work leads to some important constraints. For instance, by comparing the eclipse depth, $\delta_{ec}$, and the phase variation amplitude, $A_{ill}$, we unveiled  nightside emission for Kepler-10b and Kepler-78b, and possibly Kepler-407b and K2-106b. 
Assuming that the dayside cannot be colder than the nightside, we obtained a lower limit on the dayside temperature, namely, $T_{d} \gtrsim T_{n}$, which is comparable to the maximum theoretical value $T_{d, max}$ from thermal equilibrium considerations (see Sects.~5.1 and 5.2, along with related Figs.~\ref{fig: kepler-10b_GA} and \ref{fig: kepler-78b_GA}). This implies planetary temperatures hotter than $T_{d, max}$, possibly due to heat retention through greenhouse effects of a high molecular weight collisional atmosphere. 

The bulk density measured for the planets considered in this work indicates that they do not host a primary, hydrogen-dominated atmosphere, which was likely lost swiftly given the action of the intense stellar irradiation \citep[e.g.,][]{2017MNRAS_Lopez,kubyshkina2018grid}. Therefore, these planets may have quickly developed a secondary, possibly CO$_2$-dominated, atmosphere as a result of volcanic activity or magma ocean outgassing \citep[e.g.,][]{elkins-tanton2008magmaOcean}. Furthermore, if the secondary atmosphere formed while the star was still active, mass-loss driven by the intense stellar irradiation would have led to a complete escape also of this secondary atmosphere \citep[e.g.,][]{kulikov2006,tian2009}, leaving behind a magma ocean on the dayside that would release heavy metals into an exosphere through non-thermal processes such as sputtering \citep[e.g.,][]{pfleger2015,vidotto2018}. However, an exosphere would not be able to redistribute heat because of its non-collisional nature and the magma ocean would be present only on the dayside, leaving a cold nightside \citep{leger2011extreme}.

Kepler-78b orbits a young star \citep{sanchis2013transits}. It is therefore possible that this planet still hosts part of the CO$_2$-dominated atmosphere released following the loss of the primary hydrogen-dominated atmosphere. Indeed, the youth of the host star and the small orbital distance pose the condition for the presence of induction heating in the planetary interior \citep{kislyakova2017}, which would significantly strengthen surface volcanism and outgassing counteracting escape \citep{kislyakova2018,kislyakova2020induction-aa}. If this secondary atmosphere is dense enough to be collisional, then heat could be carried from the day to the nightside. Furthermore, Kepler-78b orbits inside the Alfv\'en radius of the star \citep{strugarek2019}, thus powering magnetic star-planet interactions that may provide further energy heating up the planetary atmosphere.

In contrast, Kepler-10b orbits an old star \citep{Fogtman-Schultz2014ApJ} suggesting that the secondary atmosphere that was built up initially may have already been lost and that neither induction heating nor magnetic star-planet interaction would be present to support a collisional atmosphere against escape. However, a further secondary, CO$_2$-dominated atmosphere may have built up over time from outgassing of the magma ocean present on the dayside as a result of the decreasing strength of mass loss with time due to the decreasing amount of high-energy radiation emitted by late-type stars with increasing age.

The scenario of magma-ocean planets with no heat redistribution as described by \citet{leger2011extreme} may apply to both K2-141b and K2-131b. Indeed, for both planets we found fully consistent values of $\delta_{ec}$ and $A_{ill}$,  hence, there is no evidence for a significant nightside emission. For Kepler-407b, K2-106b, K2-229b, and K2-312b, the great uncertainties on $\delta_{ec}$ and $A_{ill}$ prevent us from deriving useful constraints on possible nightside emissions.

As mentioned before, solely optical photometry does not allow us to precisely measure geometric albedos. It is still debated whether the albedos of USP small planets should be prevalently high or low. High albedos ($A_{g} > 0.2$) would require clouds of substantial reflective molecules in the secondary atmosphere \citep{Demory2014ApJ}. Alternatively, for near-airless planets, high albedos could be the result of specular reflections from the moderately wavy lava surfaces made of metallic species such as iron oxides \citep{Darius2021Icar}. However, low albedos ($A_{g} \lesssim 0.1$) have also been predicted for lava-ocean planets under a different theoretical framework by \citet{EssackZahra2020ApJ}. In this case, the occultation signal would be mostly due to high thermal emission (e.g. Figs. \ref{fig: k2-141b_GA}, \ref{fig: k2-131b_GA}). Moreover, we found that the corresponding brightness temperatures for very low albedos and no heat redistribution \citep{Sheets2017AJ} might be even higher than the maximum theoretical estimates. This would imply additional heat sources such as internal tidal or magnetic heating \citep[e.g.][]{Lanza2021}.

Constraining the Bond albedo ($A_{B}$) and the circulation efficiency ($\epsilon$) requires a couple of assumptions, specifically: a scattering relationship between the Bond albedo and the geometric albedo, which we assumed isotropic in this work, and thermal equilibrium between the received stellar irradiation and the planet emission. Future infrared observations, for instance, with the forthcoming James Webb Space Telescope \citep{2009PASP..121..952D_JWST}, should permit  the degeneracy between the reflected and thermally emitted light to be broken and could also provide more precise $A_{g}$, $T_{d}$, and $\epsilon$ estimates. These values, in turn, will yield valuable information on the surface or atmospheric properties of USP small planets and will greatly help in understanding the nature of these extreme worlds.

\begin{acknowledgement}
We acknowledge the computing centre of INAF - Osservatorio Astrofisico di Catania, under the coordination of the CHIPP project, for the availability of computing resources and support. We acknowledge financial contribution from the agreement ASI-INAF n.2018-16-HH. This research has made use of the NASA Exoplanet Archive, which is operated by the California Institute of Technology, under contract with the National Aeronautics and Space Administration under the Exoplanet Exploration Program. This research has also made use of publicly available Kepler data via the MAST archive and therefore, we acknowledge the funding for the \textit{Kepler} space mission provided by the NASA Science Mission Directorate and Space Telescope Science Institute (STScI) for maintaining the archive. 
\end{acknowledgement}

\bibliographystyle{aa} 
\bibliography{aanda.bib} 


\begin{appendix}

\section{Posterior distribution of the model parameters}

 \begin{figure*}
  \centering
  \includegraphics[width=0.9\linewidth]{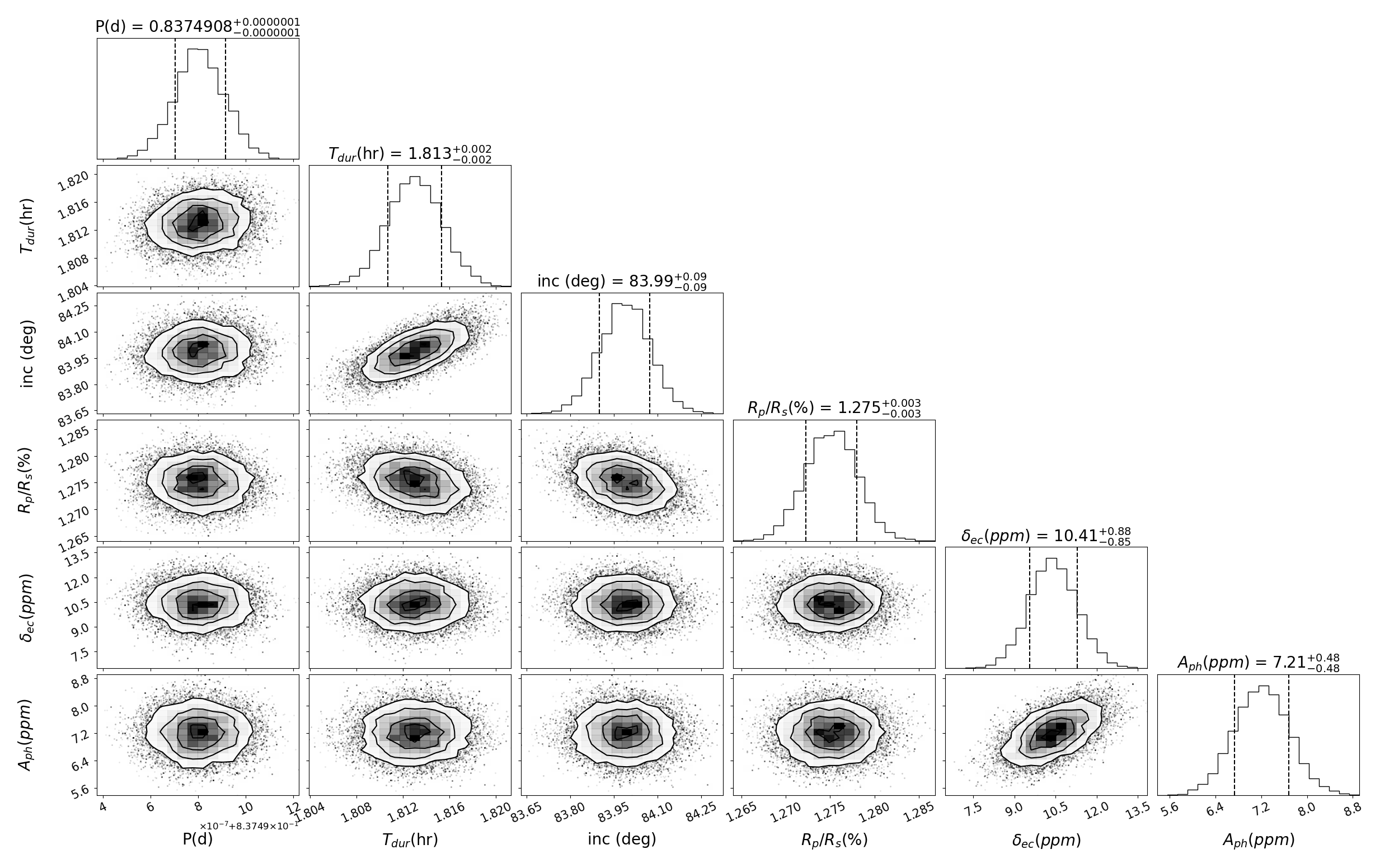}
  \caption{Posterior distribution of the best-fit model parameter of Kepler-10b.}
  \label{fig: DEMC-Kepler-10b}
 \end{figure*}

 \begin{figure*}
  \centering
   \includegraphics[width=0.9\linewidth]{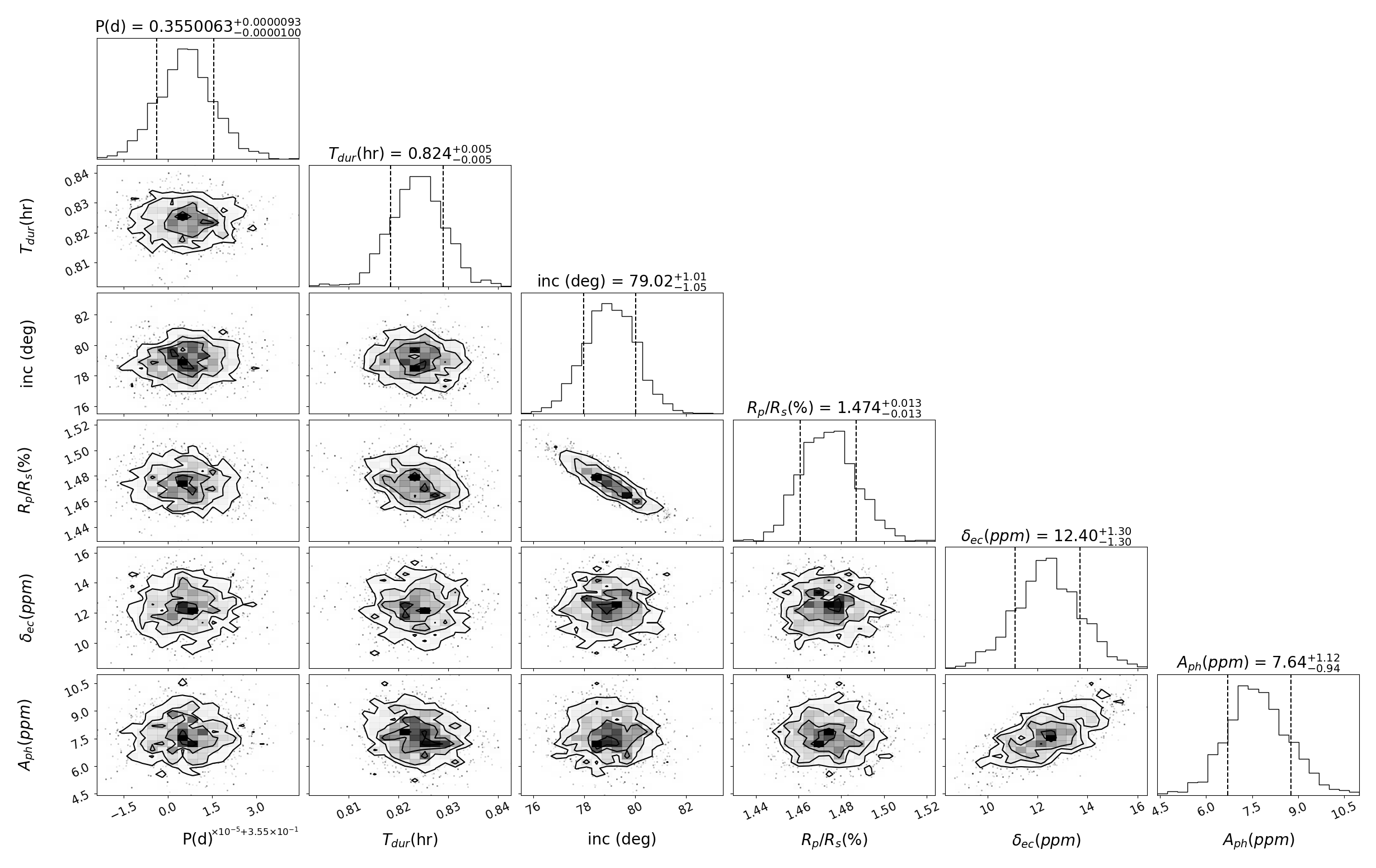}
   \caption{Posterior distribution of the best-fit model parameter of Kepler-78b.}
   \label{fig: DEMC-Kepler-78b}
 \end{figure*}

\begin{figure*}
\centering
    \includegraphics[width=0.9\linewidth]{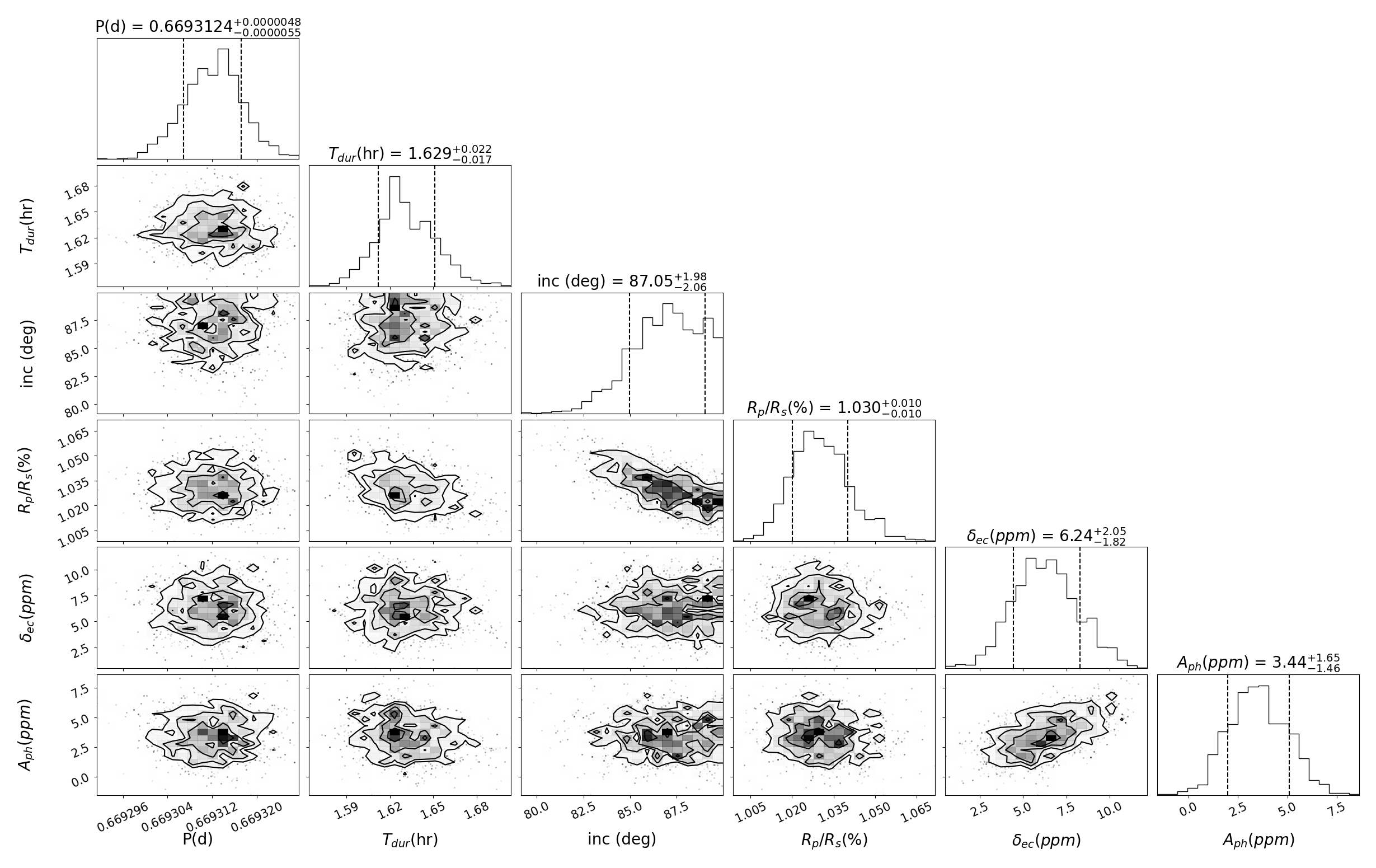}
    \hspace{1.0cm}
    \caption{Posterior distribution of the best-fit model parameter of Kepler-407b.}
    \label{fig: DEMC-Kepler-407b}
\end{figure*}

\begin{figure*}
\centering
    \includegraphics[width=0.9\linewidth]{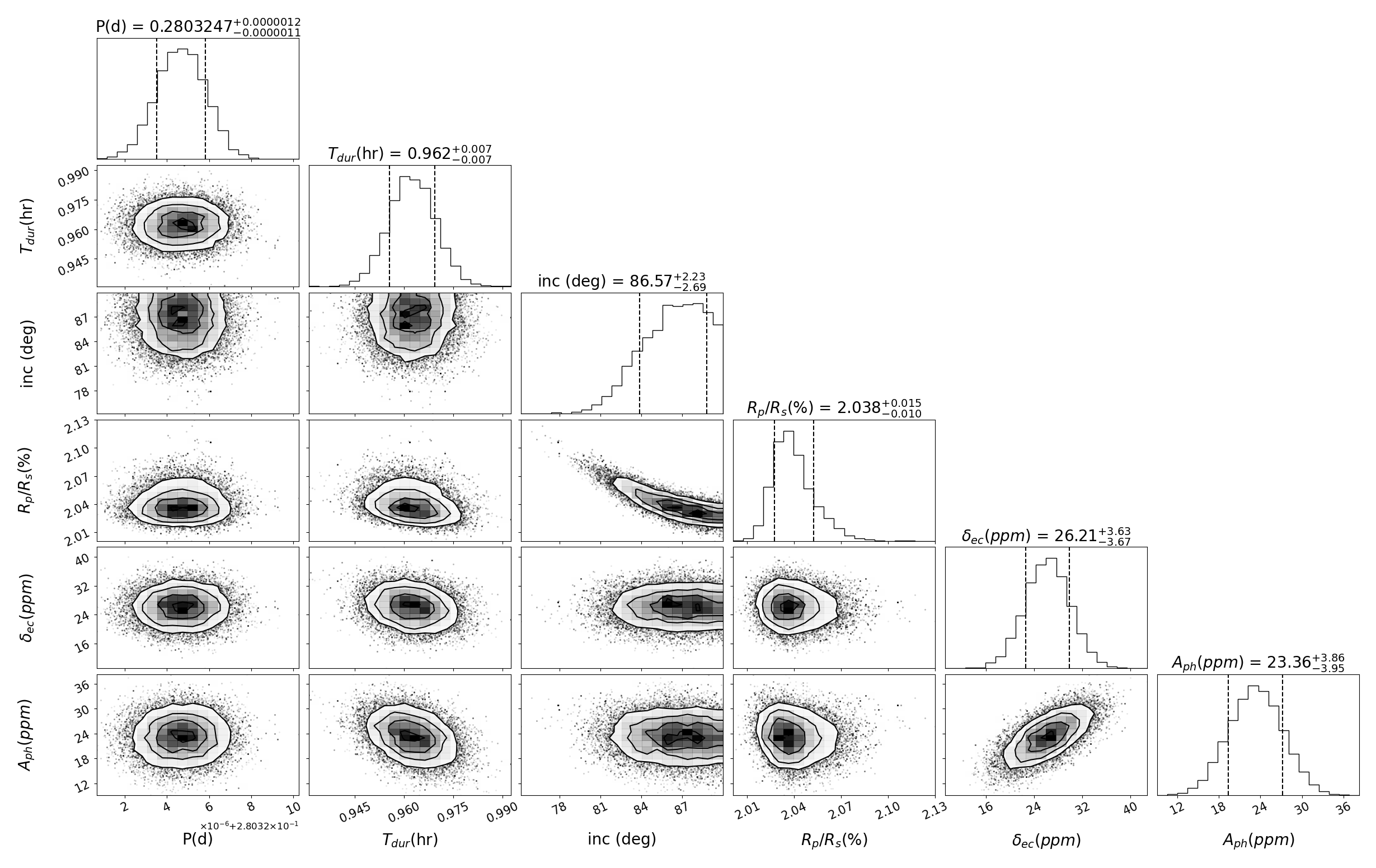}
    \hspace{1.0cm}
    \caption{Posterior distribution of the best-fit model parameter of K2-141b.}
    \label{fig: DEMC-K2-141b}
\end{figure*}

\begin{figure*}
   \centering
        \includegraphics[width=0.9\linewidth]{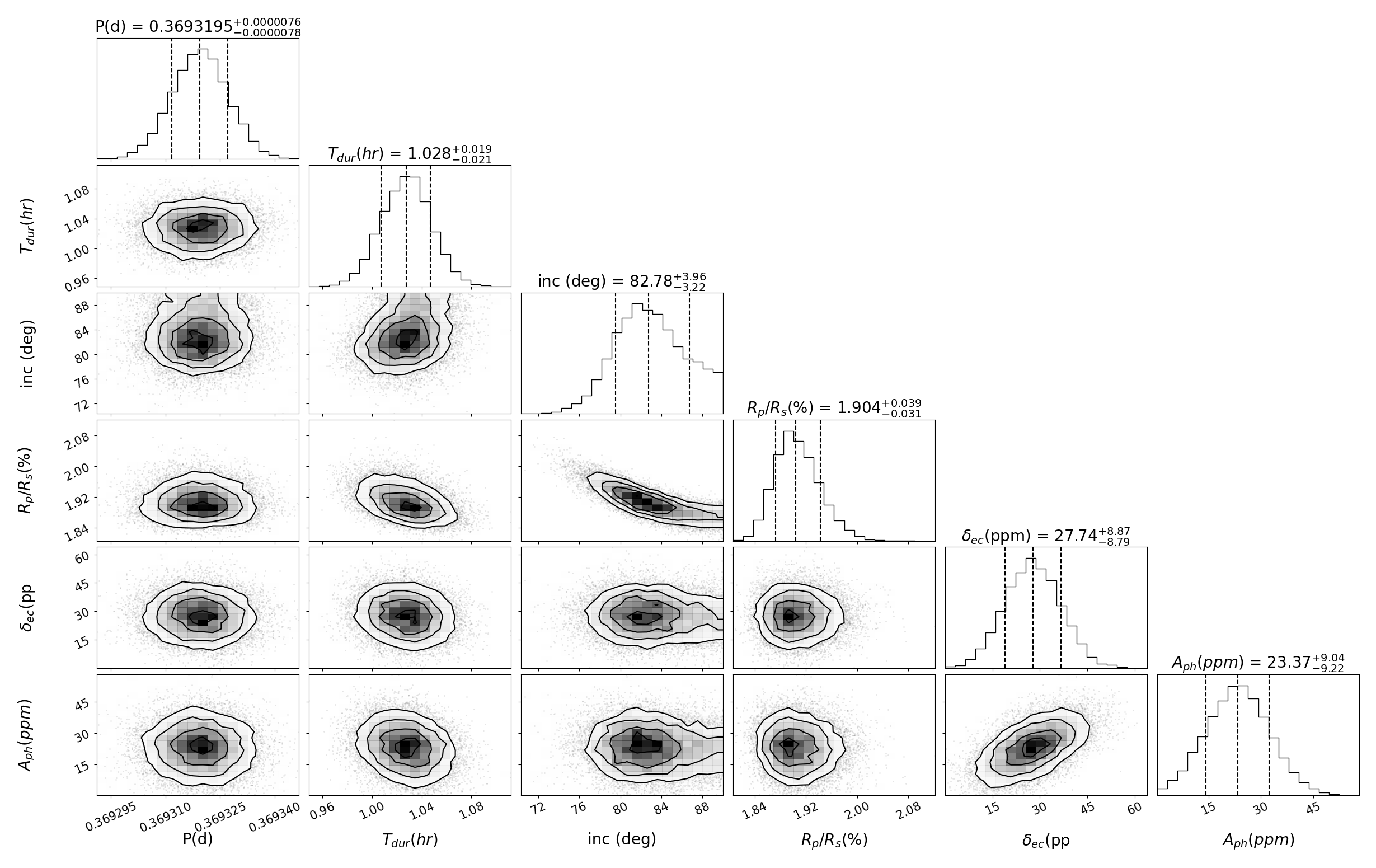}
        \hspace{1.0cm}
        \caption{Posterior distribution of the best-fit model parameter of K2-131b.}
        \label{fig: DEMC-K2-131b}
\end{figure*}

\begin{figure*}
\centering
    \includegraphics[width=0.95\linewidth]{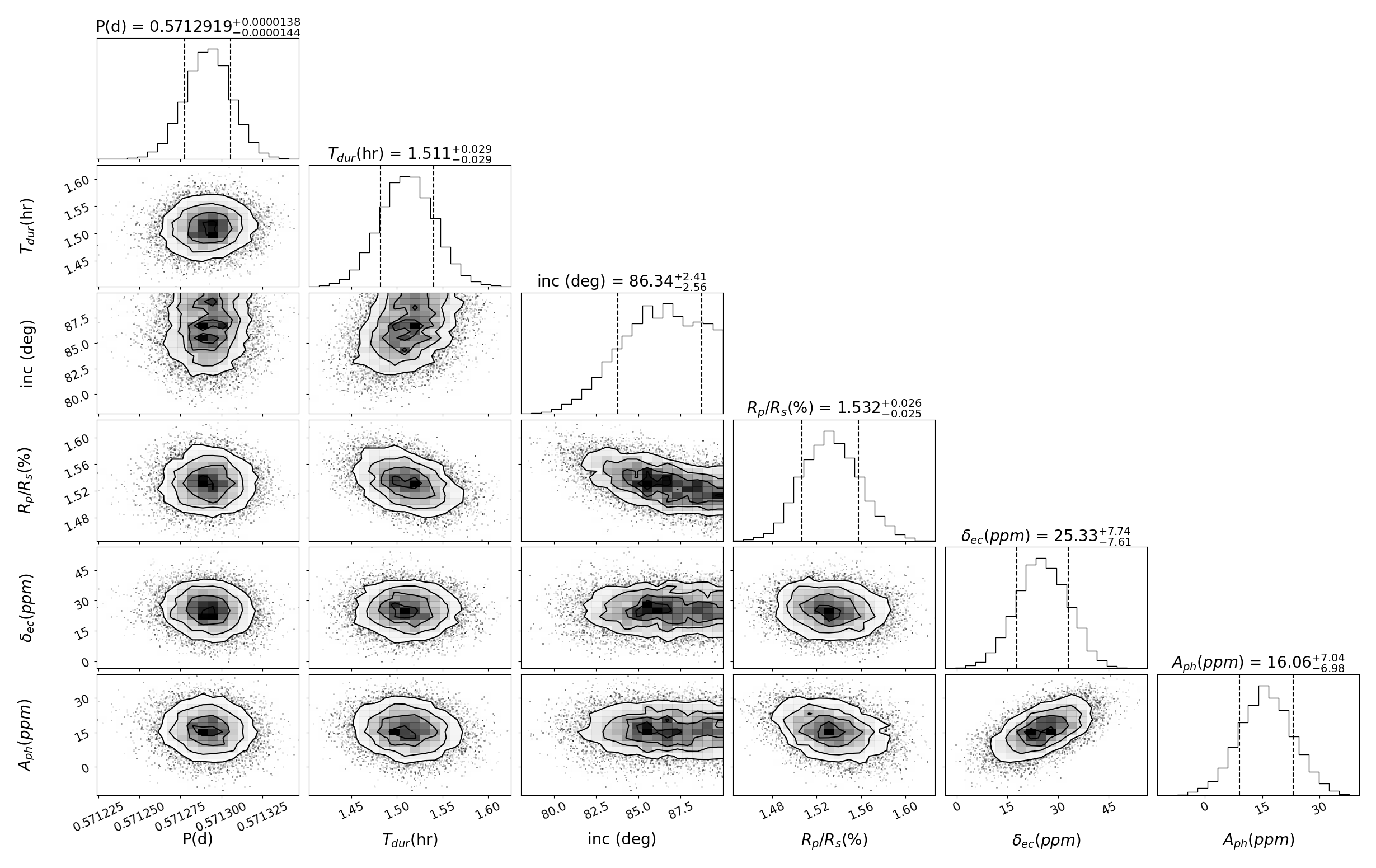}
    \caption{Posterior distribution of the best-fit model parameter of K2-106b.}
    \label{fig: DEMC-K2-106b}
\end{figure*}

\begin{figure*}
\centering
    \includegraphics[width=0.9\linewidth]{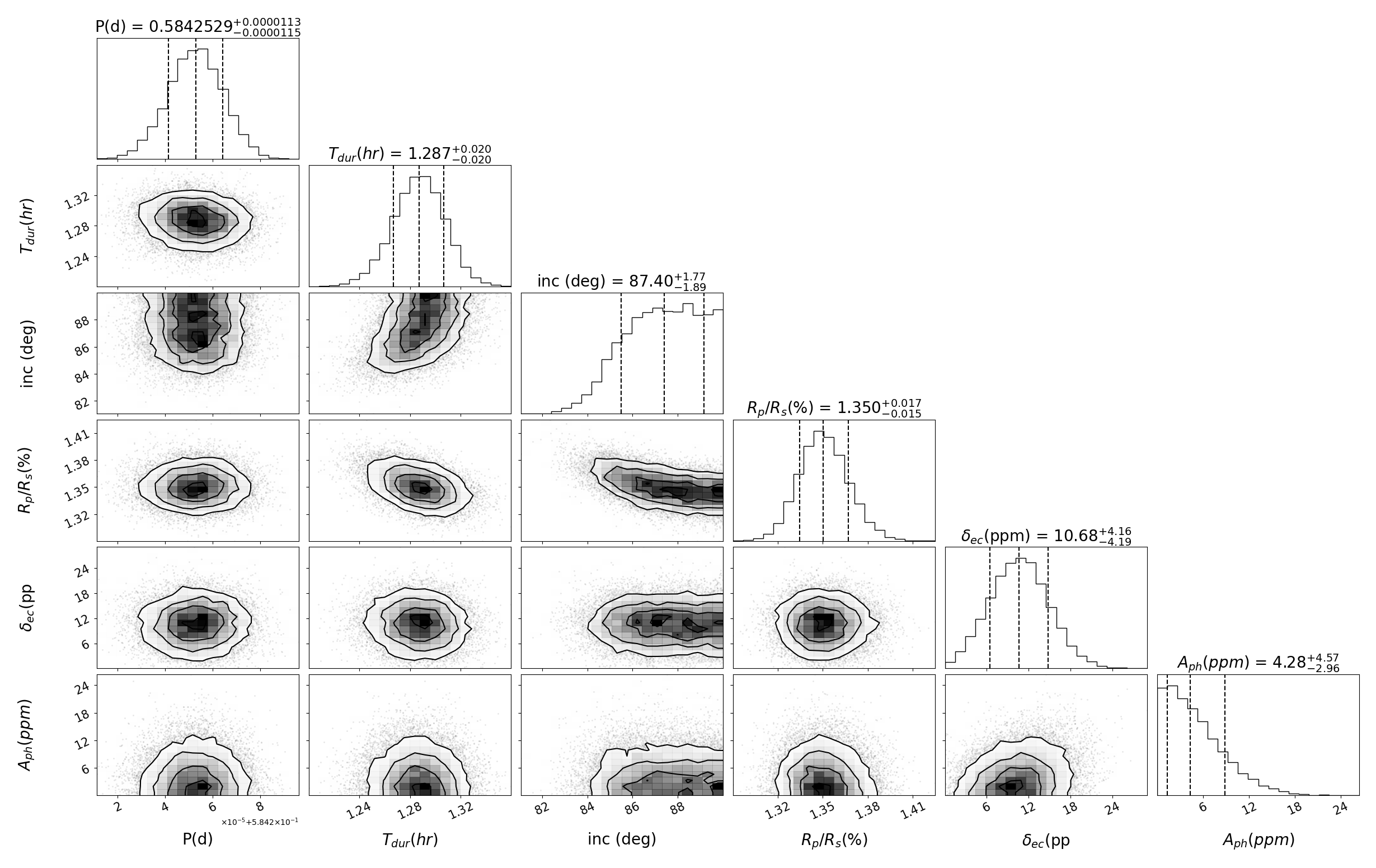}
    \hspace{1.0cm}
    \caption{Posterior distribution of the best-fit model parameter of K2-229b.}
    \label{fig: DEMC-K2-229b}
\end{figure*}

    \begin{figure*}
    \centering
        \includegraphics[width=0.9\linewidth]{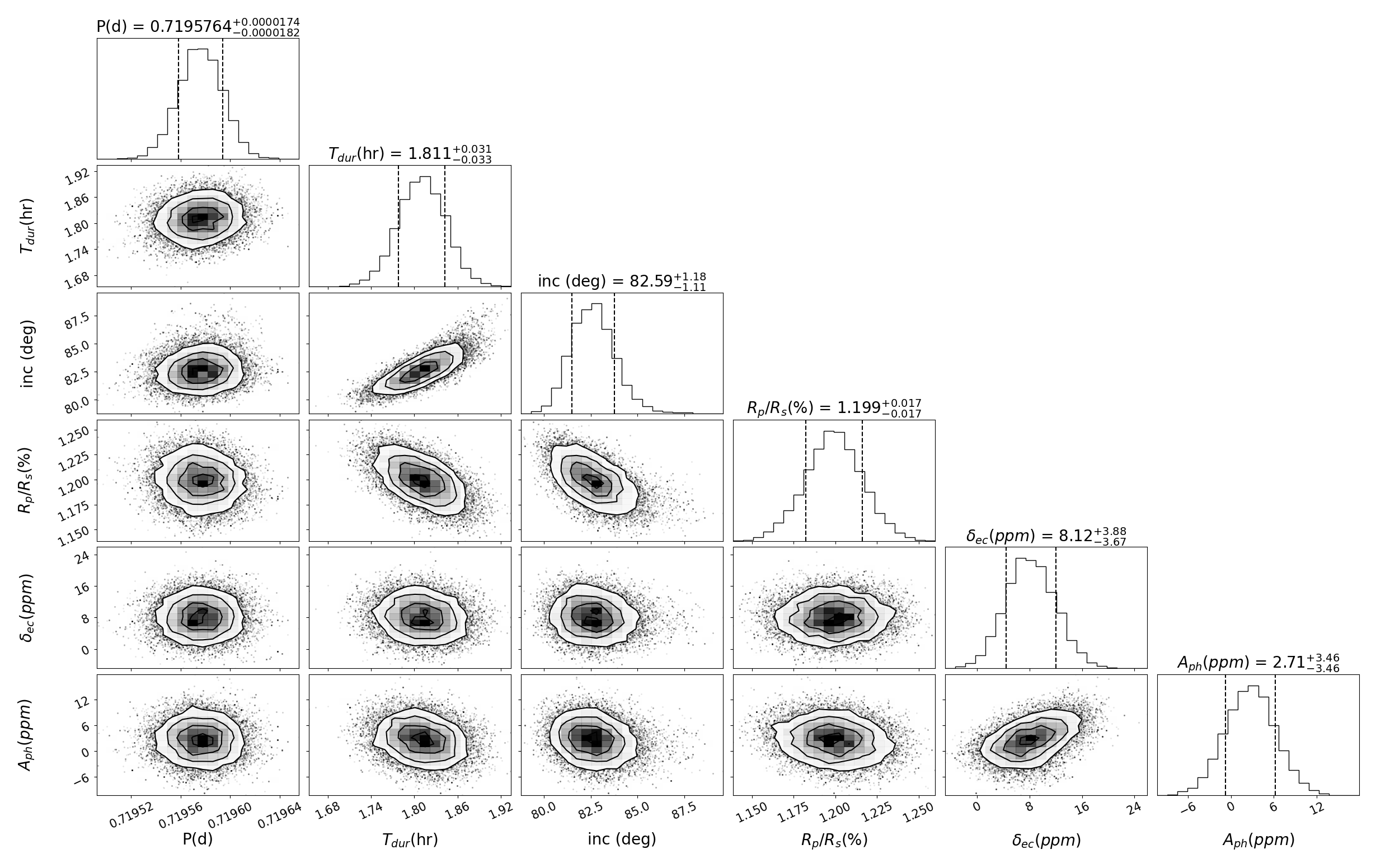}
        \hspace{1.0cm}
        \caption{Posterior distribution of the best-fit model parameter of K2-312b.}
        \label{fig: DEMC-K2_312b}
    \end{figure*}

\begin{landscape}
  \begin{figure}
  \centering
  \includegraphics[width=0.9\linewidth]{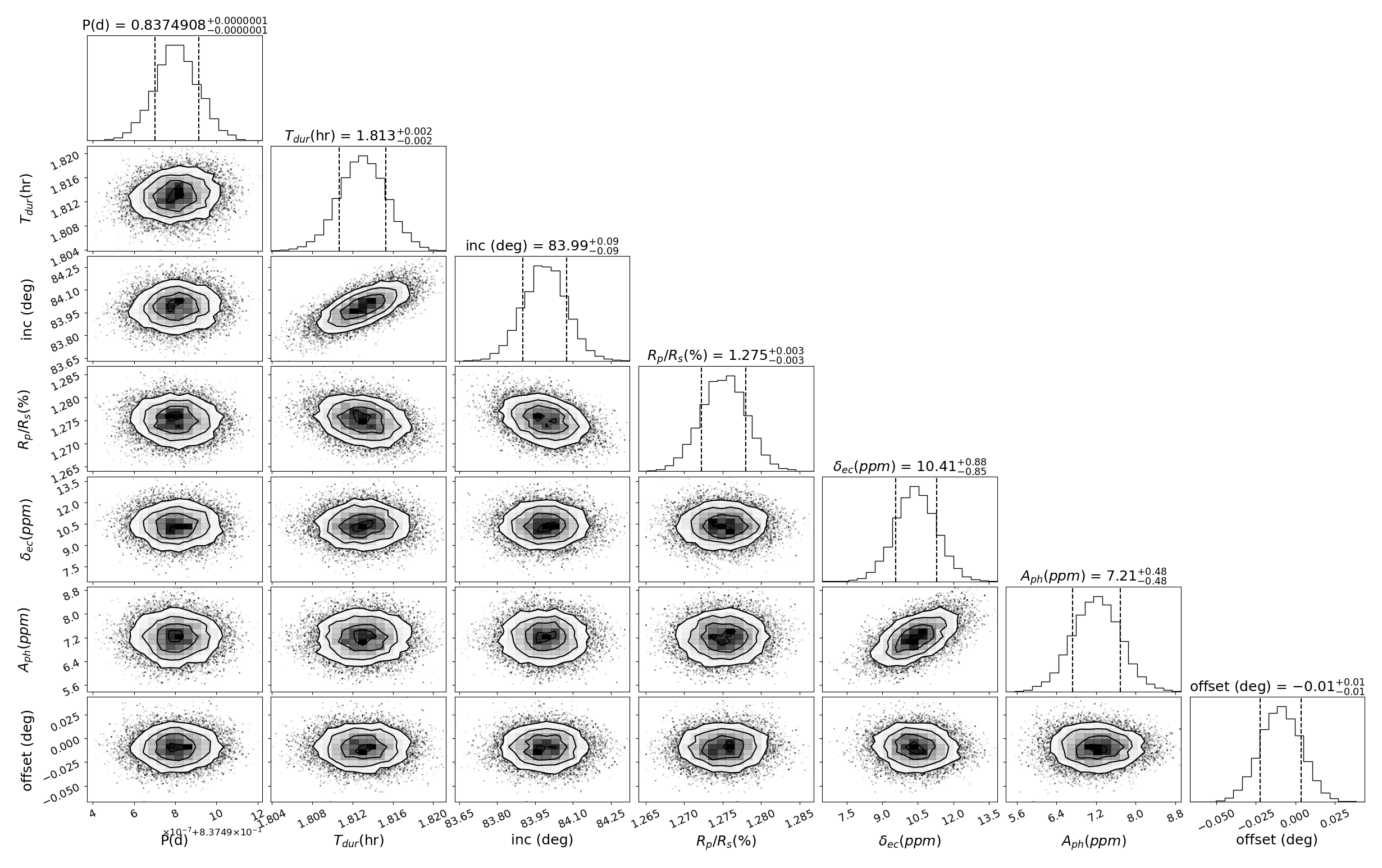}
  \caption{Posterior distribution of the best-fit model parameter of Kepler-10b including phase offset as a free parameter. Posteriors for the phase offset is consistent with zero.}
  \label{fig: DEMC-Kepler-offset}
 \end{figure}
\end{landscape}

\clearpage
\section{Detrending of stellar variability using Gaussian processes}

\begin{figure*}
\includegraphics[width=0.6\textwidth]{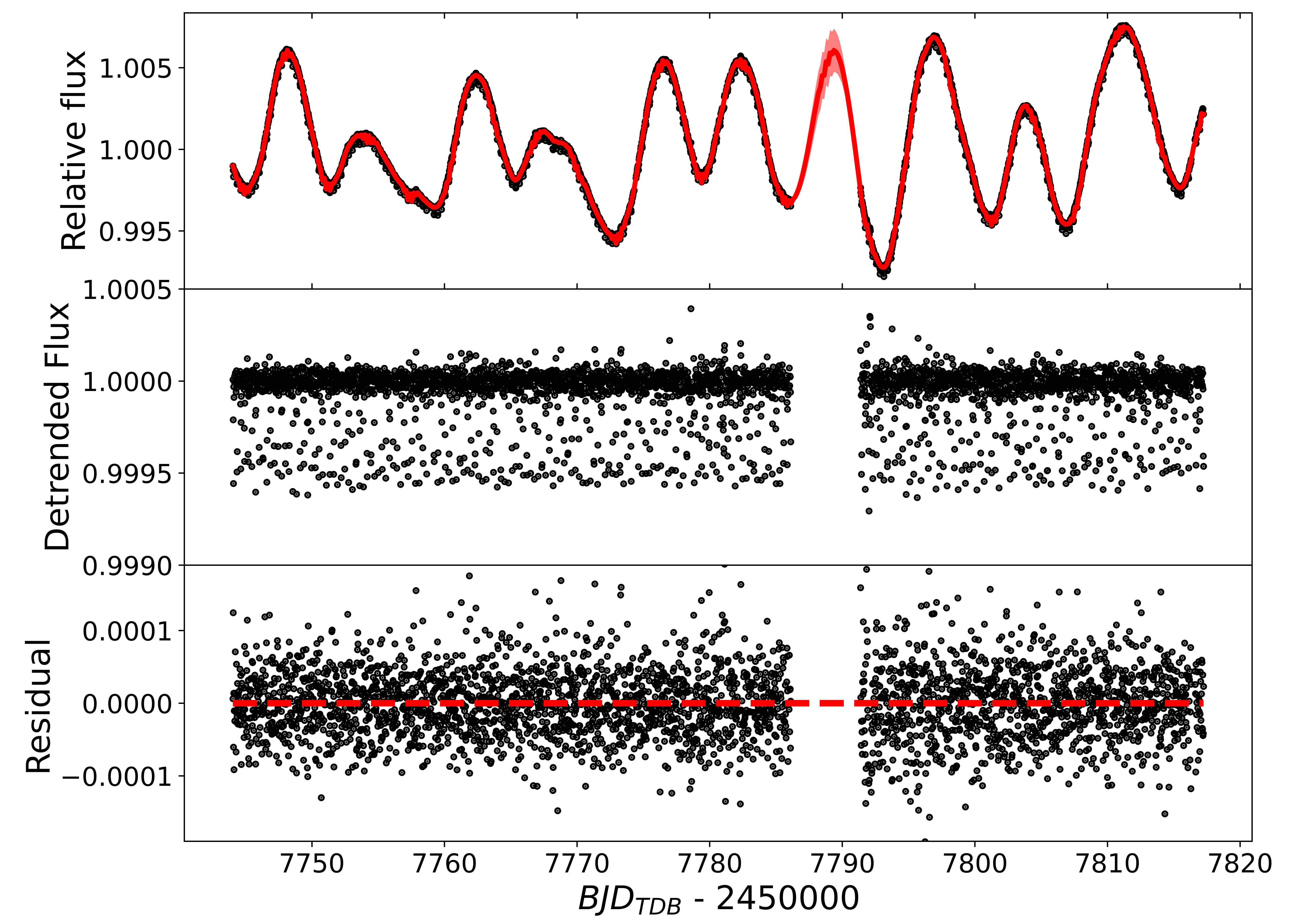}
\hspace{1.0cm}
\caption{Light curve detrending using Gaussian processes. Top: K2-141 light curve after removing the transits of planet c and with the GP model in red. Middle: Light curve minus the GP model. Bottom: residual of the combined GP + transit, eclipse and the phase variation best-fit model.}
\label{fig: K2-141_GP}
\end{figure*}

\begin{figure*}
\includegraphics[width=0.6\textwidth]{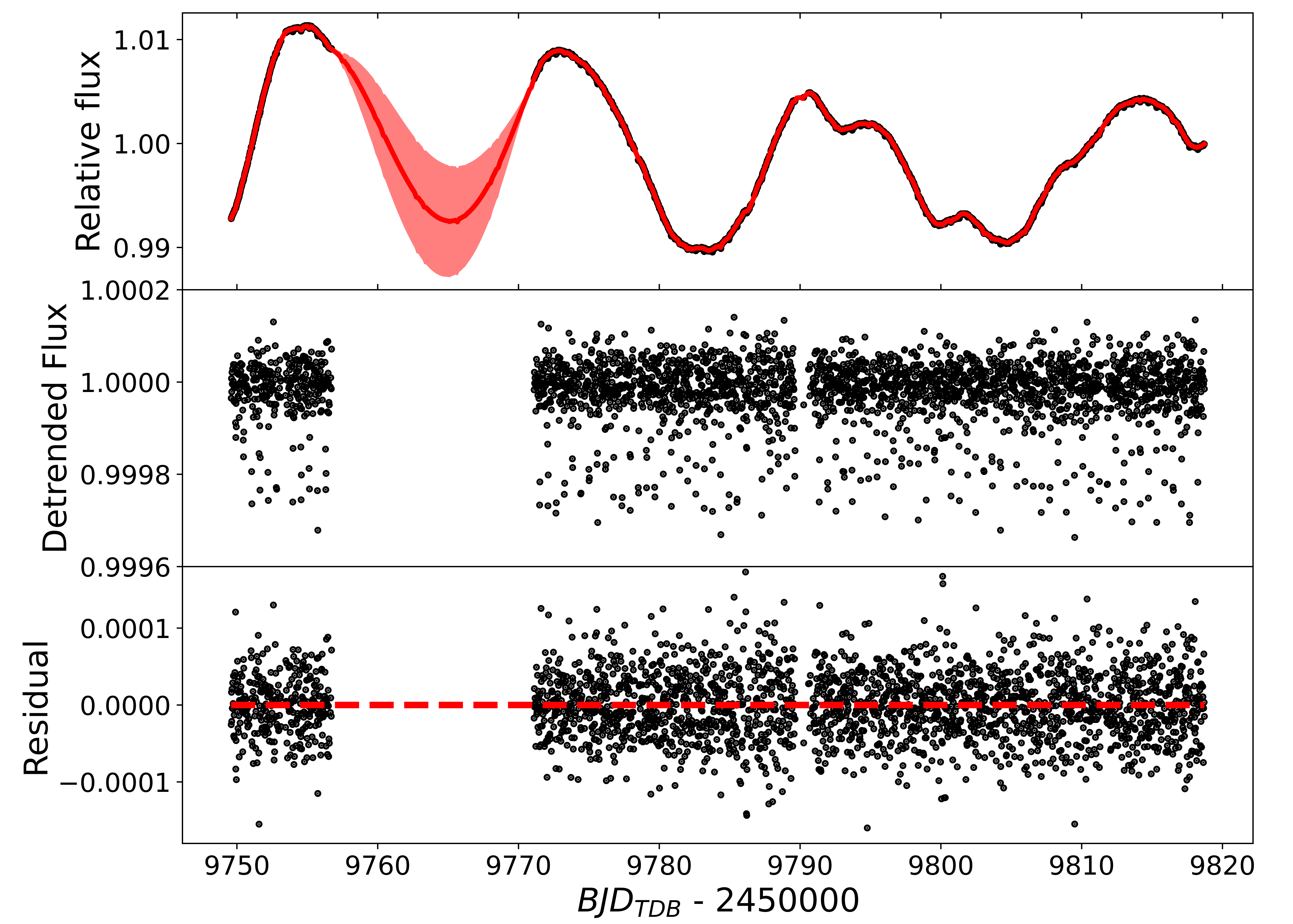}
\hspace{1.0cm}
\caption{Light curve detrending of K2-229 using Gaussian processes after removing the transits of K2-229c and K2-229d (ref. Fig. \ref{fig: K2-141_GP})}
\label{fig: K2-229_GP}
\end{figure*}

\end{appendix}

\end{document}